\documentclass [11pt,a4paper]{article}
\usepackage{jcappub}

\usepackage{amsmath}
\usepackage{amsfonts}
\usepackage{graphicx}
\usepackage{url}
\usepackage{here}

\def\be{\begin{equation}}
\def\ee{\end{equation}}
\def\ba{\begin{eqnarray}}
\def\ea{\end{eqnarray}}
\def\bs{\begin{subequations}}
\def\es{\end{subequations}}
\def\com{\color{magenta}}
\def\cob{\color{blue}}
\newcommand{\Eq}[1]{(\ref{#1})}

\newcommand{\eff}{{\text{\tiny eff}}}

\def\rme{e}
\def\rmd{d}
\def\rmi{i}

\newcommand{\oarX}[1]{\href{http://arxiv.org/abs/#1}{{\ttfamily\com #1}}}
\newcommand{\arX}[1]{\href{http://arxiv.org/abs/#1}{{\ttfamily\com arXiv:#1}}}
\newcommand{\doij}[5]{\href{http://dx.doi.org/#1}{\cob {\it #2} {\bf #3} (#5) #4}}
\newcommand{\doin}[6]{\href{http://dx.doi.org/#1}{\cob {\it #2} {\bf #3 #4} (#6) #5}}
\newcommand{\doinn}[5]{\href{http://dx.doi.org/#1}{\cob {\it #2} {\bf #3} (#5) #4}}

\newcommand{\tia}[1]{\textit{#1},}

\begin{document}

\title{Strong Planck constraints on braneworld and non-commutative inflation} 

\author[a]{Gianluca Calcagni,}
\affiliation[a]{Instituto de Estructura de la Materia, CSIC, Serrano 121, 28006 Madrid, Spain}
\emailAdd{calcagni@iem.cfmac.csic.es}

\author[b]{Sachiko Kuroyanagi,}
\affiliation[b]{Department of Physics, Faculty of Science, Tokyo University of Science, 
1-3, Kagurazaka, Shinjuku-ku, Tokyo 162-8601, Japan}
\emailAdd{skuro@rs.tus.ac.jp}

\author[b]{Junko Ohashi,}
\emailAdd{j1211703@ed.tus.ac.jp}

\author[b]{Shinji Tsujikawa} 
\emailAdd{shinji@rs.kagu.tus.ac.jp}

\abstract{
We place observational likelihood constraints on braneworld and non-commutative inflation for a number of inflaton potentials, using Planck, WMAP polarization and BAO data. Both braneworld and non-commutative scenarios of the kind considered here are limited by the most recent data even more severely than standard general-relativity models. 
At more than 95\,\% confidence level, the monomial potential $V(\phi) \propto \phi^p$ is ruled out for $p \geq 2$ in the Randall--Sundrum (RS) braneworld cosmology and, for $p>0$, also in the high-curvature limit of the Gauss--Bonnet (GB) braneworld and in the infrared limit of non-commutative inflation, due to a large scalar spectral index. Some parameter values for natural inflation, small-varying inflaton models and Starobinsky inflation are allowed in all scenarios, although some tuning is required for natural inflation in a non-commutative spacetime.
}




\maketitle

\section{Introduction}

Experimental cosmology is living its golden age. Observations of the Cosmic Microwave Background (CMB) have been approaching the ideal precision, starting with WMAP \cite{WMAP1,WMAP9} and following with the first achievements of the Planck satellite \cite{Planck1}. 
This permitted to constrain a variety of inflationary scenarios in an unprecedented 
way \cite{Planck2,Martin,Ma,Suyama,Naka,Stein,Kuro,Burgess,Bar} and to partly remove the huge model degeneracy surrounding the history of the early universe. An increasing number of theoretical proposals has been advanced to describe new physics beyond both the Standard Model and general relativity (GR), and it is becoming more and more important to appeal to experiments in order to focus one's attention to the most promising candidates.

The purpose of this paper is to take advantage of modern data and revisit some high-energy models which thrived in the late 1990s and early 2000s: the braneworld (consult \cite{MaK} for a review) and Brandenberger--Ho non-commutative inflation \cite{BH}. The first (section \ref{branesec}) is a class of string-inspired models where the observer lives in a four-dimensional manifold (the brane) embedded in five non-compact dimensions. The gravitational dynamics induced on the brane is modified with respect to standard cosmology, which leads to deviations from the inflationary predictions of general relativity. We will focus on two specific braneworld models, Randall--Sundrum (RS, \cite{RSa,RSb}) and Gauss--Bonnet (GB, \cite{KKL1,KKL2}). The non-commutative scenario of \cite{BH} and its developments \cite{HL1,Fukuma,TMB,HL2,HL3,KLM,KLLM,cai04,Cal4,Cal5} (section \ref{noncomsec}) assume, in accordance with some quantum-gravity approaches, that spacetime is not an ordinary manifold but possesses a non-commutative structure, determined by a fundamental length scale and an intrinsic uncertainty relation on space and time interval measurements. 

By now, it is clear that standard inflation with large-field potentials is under observational pressure, while small-varying inflaton models and potentials motivated by string theory or higher-curvature actions are favored \cite{Planck2,Kuro}. In all these cases, the dynamical cosmological equations are the same of general relativity, the only difference being in the choice of the potentials. 
The question is whether braneworld and non-commutative scenarios, where the dynamics is modified, fare better when the same types of potentials are considered. Performing a likelihood analysis of cosmological data (section \ref{setup}), we shall see in sections \ref{braneconsec} and \ref{branenoncomsec} that this is not the case. Large-field potentials are even less viable than in standard cosmology, while small-varying inflaton models and Starobinsky inflation still survive but in a narrower parameter space. Overall,  braneworld and non-commutative inflation do not present any advantage or characteristic signature with respect to standard cosmology, and in large portions of their parameter space they are actually excluded.

This drastic conclusion was not possible ten years ago, when constraints with the WMAP1 
data \cite{WMAP1} were imposed following the same method \cite{CT04}.\footnote{In section \ref{setup} we will comment on studies which appeared between \cite{CT04} and the present paper. 
None of them performed a likelihood analysis.} The area in the $(n_{\rm s},r)$ plane enclosed in likelihood contours was large enough to include most of the theoretical points of these scenarios, apart from the exponential potential in some cases. The resulting non-committal conclusion was that braneworld and non-commutative models could be discriminated from standard inflation by further data and near-future experiments. The shrinking of the likelihood areas has now come to the exciting point where we are in that `near future' and we can finally place the desired constraints.\footnote{This paper does not reflect the BICEP2 data of the $B$-mode polarization that appeared 5 months after the initial submission.}

\section{Braneworld inflation}\label{branesec}

In braneworld scenarios, the universe and its observers live in a (3+1)-dimensional manifold (a brane) embedded in a larger non-compact spacetime (the bulk). Such configuration, with a five-dimensional bulk, arises as the low-energy limit of the Calabi--Yau compactification of six dimensions in M theory \cite{LOSW,LOW1}. Assuming a Friedmann--Lema\^{i}tre--Robertson--Walker (FLRW) background on the brane and that all the matter is confined therein (the 5D energy-momentum tensor is $T^{\mu\nu}\propto\delta(y_b)\,{\rm diag}(\rho,-p,-p,-p,0)$, where $y_b$ is the brane position along the extra direction $y$), if the 5D gravitational action is the Einstein--Hilbert one (Randall--Sundrum model) one can show that the Friedmann equation acquires a quadratic correction \cite{BDL,BDEL,Shiromizu},
\be\label{rsf}
H^2 = \frac{\kappa_4^2}{6 \lambda} \rho (2 \lambda+\rho)+\frac{\cal E}{a^4}\,,
\ee
where $H=\dot a/a$ is the Hubble parameter, $a$ is the scale factor, a dot represents a derivative with respect to synchronous time $t$, $\kappa_4^2:= 8\pi/m_{\rm Pl}^2$ includes the four-dimensional Planck mass $m_{\rm Pl} \approx 10^{19}\,\text{GeV}$, $\lambda$ is the brane tension and ${\cal E}=$const.\ is a dark radiation term. The RS model can be viewed as an intermediate scenario between the standard 4D (low-energy) evolution, $H^2 \propto \rho$ and a braneworld scenario where higher-order curvature corrections, arising in the heterotic string \cite{GrS}, are included in the 5D action:
\be
S =\frac{1}{2\kappa_5^2} \int \rmd^5x\sqrt{-g_5}\left[R-2\Lambda_5+\alpha_{\rm GB}\left(R^2-4R_{\mu\nu}R^{\mu\nu}+R_{\mu\nu\rho\sigma}R^{\mu\nu\rho\sigma}\right)\right]+S_\text{\tiny matter}\,.
\ee
Here, $\kappa_5$ is the 5D gravitational coupling, $g_5$ is the determinant of the 5D metric, 
$R$ is the 5D Ricci scalar,
$\Lambda_5<0$ is the bulk cosmological constant and $\alpha_{\rm GB}=1/(8g_{\rm s}^2)>0$ is the coupling of the Gauss--Bonnet term, where $g_{\rm s}$ is the string energy scale. We omitted a boundary term in the action. The effective Friedmann equation on the brane is \cite{CD,dav03,GW,Torii}
\be\label{gabo}
H^2=\frac{c_+ + c_- -2}{8\alpha_{\rm GB}}\,,
\ee
where $H$ is the Hubble parameter and, defining $\delta_0^{-1}:=\sqrt{\alpha_{\rm GB}/2}\,\kappa_5^2$,
\be
c_\pm = \left[\sqrt{\left(1+4\alpha_{\rm GB}\Lambda_5/3\right)^{3/2}+\left(\delta/\delta_0\right)^2} \pm \delta/\delta_0\right]^{2/3},
\ee
$\delta=\rho+\lambda$ being the matter energy density decomposed into a matter contribution plus the brane tension $\lambda$. Expanding Eq.\ (\ref{gabo}) to quadratic order in $\delta$, one recovers the Friedmann equation \Eq{rsf} of the RS scenario with vanishing 4D cosmological constant, provided some relations between the couplings of the model are satisfied. Therefore, one can now recognize three energy regimes: 
\ba
& &H^2\approx \left(\frac{\kappa_5^2}{16\alpha_{\rm GB}}\right)^{2/3}\rho^{2/3} \quad 
{\rm when} \quad \delta/\delta_0 \gg 1~~({\rm GB~regime}),\\
& &H^2 \approx \frac{\kappa_4^2}{6\lambda}\rho^2 \qquad \qquad~~
{\rm when} \quad \lambda/\delta_0 \ll \delta/\delta_0 \ll 1~~({\rm RS~regime}),\\
& &H^2 \approx \frac{\kappa_4^2}{3}\rho \qquad \qquad \qquad~
{\rm when} \quad \rho/\delta_0 \ll \delta/\delta_0 \ll 1~~({\rm GR~regime}).
\ea

An economic way to treat all these regimes analytically is the patch formalism \cite{Cal5,Cal3}, where one assumes the effective Friedmann equation
\be 
\label{Hubble}
H^2=\beta_q^2 \rho^q\,,
\ee
and $\beta_q$ and $q$ are constants. In different energy regimes and time intervals (the patches), $q$ acquires different values:
\ba 
q = \begin{cases}
1 & ({\rm GR}), \\
2 &  ({\rm RS})\,,\\
2/3 & ({\rm GB})\,.
\end{cases}
\ea

We focus on the case in which a minimally coupled scalar field 
$\phi$ is confined on the 3-brane.
On the homogeneous and isotropic background, 
the energy density of $\phi$ is given by  
\be \label{rho}
\rho=\frac12 \dot{\phi}^2+V(\phi)\,,
\ee
where $V(\phi)$ is the potential of $\phi$. 
The inflaton satisfies the following equation of motion:
\be 
\label{phieq}
\ddot{\phi}+3H\dot{\phi}+V_{,\phi}(\phi)=0\,,
\ee
where $V_{,\phi}=dV/d\phi$.

The slow-roll (SR) parameters associated with the inflaton $\phi$ are 
\be
\epsilon_{\phi}:=-\frac{\dot{H}}{H^2}\,,\qquad
\eta_{\phi}:=-\frac{\ddot{\phi}}{H \dot{\phi}}\,,\qquad
\xi_{\phi}^2:=\frac{1}{H^2} \left( \frac{\ddot{\phi}}{\dot{\phi}}
\right)^{\cdot}\,.
\label{Hxi}
\ee
We also introduce the horizon-flow (HF) 
parameters \cite{Sch,Leach} 
\be
\epsilon_0:=\frac{H_{\rm inf}}{H}\,,\qquad
\epsilon_{j+1}:=\frac{ d \ln |\epsilon_j|}{dN}\,,\qquad
j \ge 0\,,
\label{hflow}
\ee
where $H_{\rm inf}$ is the Hubble rate at some chosen time and 
$N := \ln (a/a_{\rm i})$ is the number of e-folds; here 
$a_{i}=a(t_{\rm i})$ and $t_{\rm i}$ 
is the time at the onset of inflation. 
The evolution equation for the HF parameters is given by  
$\dot{\epsilon}_j = H\epsilon_{j}\epsilon_{j+1}$. 
The HF parameters are related to the first SR 
parameters by
\bs\ba
\epsilon_1 &=& \epsilon_\phi\,,\label{ep1}\\
\epsilon_2 &=& 2\epsilon_{\phi}/q-2\eta_{\phi}\,,\label{ep2}\\
\epsilon_2\epsilon_3 &=& 
4\epsilon_{\phi}^2/q^2-2(1+2/q)\epsilon_{\phi} \eta_{\phi}+
2\xi_{\phi}^2\,.
\ea\es

The amplitudes of scalar and tensor perturbations are 
given, respectively, 
by \cite{Maar,Langlois,Smith,TsuLiddle,Dufaux,TSM,Cal3}
\ba
\label{ampli}
{\cal P}_{\rm s} &=& 
\frac{3q\beta_q^{2/q}}{25\pi^2}
\frac{H^{4-2/q}}{2\epsilon_1}\,, \label{Ps} \\
{\cal P}_{\rm t} &=& 
\frac{48q\beta_q^{2/q}}{25\pi^2}
\frac{H^{4-2/q}}{2\zeta_q}\,,
 \label{Pt}
\ea
where 
\ba 
\zeta_q = 
\begin{cases}
1 & ({\rm GR}), \\
2/3 &  ({\rm RS})\,,\\
1 & ({\rm GB})\,.
\end{cases}
\ea
The spectra (\ref{Ps}) and (\ref{Pt}) should be evaluated when 
the modes with the physical wave number $k/a$ crossed the Hubble radius 
during inflation ($k/a=H$).
The spectral indices of scalar and tensor perturbations 
evaluated at the Hubble radius crossing are
\ba
\label{nsb}
n_{\rm s}-1  &=& \frac{d \ln {\cal P}_{\rm s}}{d\ln k}=
-(4-2/q)\epsilon_1-\epsilon_2\,, \\
n_{\rm t} &=& \frac{d \ln {\cal P}_{\rm t}}{d\ln k}=
-(4-2/q)\epsilon_1\,.
\label{ntb}
\ea
The tensor-to-scalar ratio is 
\ba
r=\frac{{\cal P}_{\rm t}}{{\cal P}_{\rm s}}=
\frac{16\epsilon_1}{\zeta_q}
=-\frac{8q}{(2q-1)\zeta_q}n_{\rm t}\,,
\label{ratiob1}
\ea
where we used Eq.~(\ref{ntb}).
The consistency relation is given by 
\ba
r&=&-8n_{\rm t} \qquad~\,({\rm GR~and~RS})\,,
\label{ratiob0}\\
r&=&-16n_{\rm t} \qquad  ({\rm GB})\,.
\label{ratiob2}
\ea
The runnings of the spectral indices, 
$\alpha_{{\rm s},{\rm t}}=d n_{{\rm s},{\rm t}}/d\ln k$, read
\ba
\alpha_{\rm s} &=& -(4-2/q)\epsilon_1 \epsilon_2-\epsilon_2 \epsilon_3\,,\\
\alpha_{\rm t} &=& -(4-2/q)\epsilon_1 \epsilon_2\,,
\ea
which are second order in the slow-roll parameters.

\section{Non-commutative inflation} \label{noncomsec}

In this section, we briefly recall the main features of non-commutative inflation \cite{BH,Cal4,Cal5}. Time and space measurements are subject to an uncertainty relation $\Delta t \Delta x_p \geq l_{\rm s}^2$, where $l_{\rm s}\equiv M_{\rm s}^{-1}$ is a fundamental length scale (possibly identifiable with the string length \cite{yon87,LY,yon00}) and $x_p$ is the physical spatial coordinate. Due to this intrinsic scale in the geometry, coordinates do not commute; an algebra preserving the maximal symmetry of the FLRW background is $[\tau,x]=\rmi l_{\rm s}^2$, where $\tau := \int a\,dt$ and $x$ is a comoving spatial coordinate. In this setting, there appears a comoving scale $k_0(\delta)$ dependent on the parameter $\delta := (M_{\rm s}/H)^2$. The space of comoving wave numbers is divided into two regions, one including small-scale perturbations generated in the ultraviolet (UV), i.e., inside the Hubble horizon ($H \ll M_{\rm s}$) and the other describing the infrared (IR), large-scale
  perturbations created outside the horizon ($H \gg M_{\rm s}$). Somewhat counterintuitively, the UV region corresponds to a weak non-commutative regime, while the IR region is characterized by strong non-commutative effects. 

The spectra of scalar and tensor perturbations turn out to be
\be 
\label{Anoncom}
{\cal P}(\delta,\,H,\,\phi)
= {\cal P}^{(c)} (H,\,\phi)\,\Sigma^2 (\delta)\,,
\ee
where ${\cal P}^{(c)}={\cal P}(\Sigma\!\!=\!\!1)$ is the amplitude in the 
commutative limit and $\Sigma(\delta)$ is a function encoding 
non-commutative effects. The same factor $\Sigma$ multiplies both the 
tensor and scalar amplitudes, so their ratio is unchanged.

Let us concentrate on the IR limit, which bears the largest non-commutative effect. The spectra (\ref{Anoncom}) are evaluated at the moment when the perturbation with comoving wave-number $k$ is generated. 
To lowest order in the HF parameters,
\be \label{dotsig}
\frac{d \ln \Sigma^2}{d \ln k}
= \sigma \epsilon_1\,,
\ee
where $\sigma = \sigma(\delta)$ is a function of $\delta$ such that 
$\dot{\sigma}=O(\epsilon_1)$. The commutative spacetime corresponds to $\sigma=0$.

Let us accommodate the effect of non-commutativity for $q=1$ in Eq.~(\ref{Hubble}).
Then, the spectral indices (\ref{nsb}) and (\ref{ntb}) are modified to 
\ba
\label{ns}
n_{\rm s}-1 &=& -(2-\sigma)\epsilon_1-\epsilon_2\,, \\
n_{\rm t} &=& -(2-\sigma)\epsilon_1\,.
\label{nt}
\ea
The tensor-to-scalar ratio reads
\be 
\label{ratio}
r=16\epsilon_1=
-\frac{16}{2-\sigma}n_{\rm t}\qquad (\sigma\neq 2)\,.
\ee
Consequently, the runnings of the spectral indices are
\ba
\alpha_{\rm s} &=& 
-(2-\sigma)\epsilon_1\epsilon_2-\sigma\bar{\sigma}
\epsilon_1^2-\epsilon_2\epsilon_3\,,
\label{alphas} \\
\alpha_{\rm t} &=& 
-(2-\sigma)\epsilon_1\epsilon_2-
\sigma\bar{\sigma}\epsilon_1^2\,,
\label{alphat}
\ea
where $\bar{\sigma} := -\dot{\sigma}/(H\sigma\epsilon_1)$.

In the infrared region, two classes of non-commutative models have been proposed \cite{BH}. In the first one (class 1), the FLRW 2-sphere is factored out in the measure $z_k$ of the effective perturbation action. The measure is thus given by the product of the commutative contribution $z$ times a $(1+1)$-dimensional correction factor. In the class-2 choice, the scale factor in the measure is everywhere substituted by an effective scale $a_\eff$ whose time dependence is smeared out by non-local effects; since $z \propto a$, then $z_k=z a_\eff/a$. Inequivalent prescriptions on the ordering of the *-product in the perturbation action further split these two classes, but in the IR limit they give almost the same predictions \cite{Cal4}. In this limit, the parameter $\sigma$ approaches the constant value $\sigma=6$ in class-1 models and $\sigma=2$ in class-2 models \cite{Cal4}.

From Eqs.~(\ref{ns})-(\ref{ratio}), it follows that 
\ba
& & n_{\rm s}-1 = 4\epsilon_1-\epsilon_2\,,\quad  \,n_{\rm t}=r/4 \qquad (\sigma=6), 
\label{nt0}\\
&& n_{\rm s}-1=-\epsilon_2\,,\qquad \quad n_{\rm t}=0 \quad \qquad(\sigma=2).
\label{nt2}
\ea
The standard GR commutative consistency relation $n_{\rm t}=-r/8$ is thus deformed by non-commutativity. 

\section{Likelihood analysis} \label{setup}

In order to place observational constraints on the inflationary models 
discussed in Secs.~\ref{branesec} and \ref{noncomsec},
the power spectra ${\cal P}_{\rm s}$ and ${\cal P}_{\rm t}$ are expanded 
around the pivot wave number $k_0$, as
\bs\label{expansion}\ba
\ln {\cal P}_{\rm s} (k) &=& \ln  {\cal P}_{\rm s} (k_0)+[n_{\rm s}(k_0)-1]x+
\frac{\alpha_{\rm s}(k_0)}{2}x^2+O(x^3), \\
\ln {\cal P}_{\rm t} (k) &=& \ln  {\cal P}_{\rm t} (k_0)+n_{\rm t}(k_0)x+
\frac{\alpha_{\rm t}(k_0)}{2}x^2+O(x^3),
\ea\es
where $x=\ln (k/k_0)$. 
For the scales relevant to the observed CMB anisotropies (the multipoles 
$2 \le \ell \lesssim 2500$), $x$ is smaller than 7. 
Since $\alpha_{{\rm s},{\rm t}}(k_0)$ are of the order of $\epsilon_j^2$, 
the third and fourth terms on the right-hand side of Eq.~(\ref{expansion}) are 
suppressed relatively to the first two terms. 
In the slow-roll expressions for $n_{\rm s,t}$ and $r$, we neglect the second-order terms, 
which are always subdominant with respect to the $O(\epsilon_j)$ parts. 
On the other hand, we retain the terms $\alpha_{\rm s,t}x^2=O(\epsilon_j^2\,x^2)$ because, although the runnings are second order in slow-roll parameters, they give rise to non-negligible effects for large $x$. 
This mixed truncation scheme is fairly standard in CMB analysis.

We choose the pivot wavenumber to be 
\be
k_0=0.05~{\rm Mpc}^{-1}\,.
\label{k0}
\ee
This is different from the value $k_0=0.002~{\rm Mpc}^{-1}$ used
by the Planck team \cite{Planck1}, but we confirm that the likelihood results are 
insensitive to the choice of $k_0$ for the scales relevant to the observed 
CMB anisotropies.

We run the CosmoMC code \cite{cosmomc,Lewis} by using 
the recent data of Planck \cite{Planck1}, WMAP polarization (WP) \cite{WMAP9}, 
Baryon Acoustic Oscillations (BAO) \cite{BAO} and high-$\ell$ 
ACT/SPT temperature data \cite{Das}. 
We use the big-bang nucleosynthesis consistency relation, by which 
the helium fraction $Y_p$ is expressed in terms of $N_{\rm eff}$ 
and the baryon fraction $\Omega_b h^2$.
The flat $\Lambda$CDM model is assumed with $N_{\rm eff} = 3.046$ 
relativistic degrees of freedom and with instant reionization.

We have six inflationary observables ${\cal P}_{\rm s}(k_0)$, $n_{\rm s}(k_0)$, 
$n_{\rm t}(k_0)$, $r(k_0)={\cal P}_{\rm t}(k_0)/{\cal P}_{\rm s}(k_0)$, 
$\alpha_{\rm s}(k_0)$, and $\alpha_{\rm t}(k_0)$ to confront with the data.
In the slow-roll framework, these reduce to four observables 
for given values of $q$ and $\sigma$. 
In both braneworld and non-commutative inflation
the scalar spectral index and the tensor-to-scalar ratio 
can be expressed as $n_{\rm s}-1=-(4-2/q-\sigma)\epsilon_1
-\epsilon_2$ and $r=16\epsilon_1/\zeta_q$, 
so that $\epsilon_1$ and $\epsilon_2$ are inverted to give
\ba
\epsilon_1 &=& 
\frac{\zeta_q}{16}r\,,\\
\epsilon_2 &=& -\frac{\zeta_q}{16} 
(4-2/q-\sigma)r+1-n_{\rm s}\,,
\ea
where we omitted the $k_0$ dependence.
Substituting these relations into Eqs.~(\ref{nt}) and (\ref{alphat}), 
$n_{\rm t}$ and $\alpha_{\rm t}$ are known.
The scalar running includes the additional parameter $\epsilon_3$, 
so we need to vary the four parameters ${\cal P}_{\rm s}$, $n_{\rm s}$, $r$ and 
$\epsilon_3$ in the likelihood analysis. The slow-roll parameter 
$\epsilon_3$ is smaller than $O(0.01)$. We have run the 
numerical code by setting the prior $\epsilon_3<0.05$ and 
found that the likelihood results are practically the same 
as those obtained for $\epsilon_3=0$.
Therefore, we set $\epsilon_3=0$ in the whole likelihood 
analysis that follows.  
By assuming $\epsilon_3=0$, we reduce the free parameters 
to only three: ${\cal P}_{\rm s}$, $n_{\rm s}$ and $r$. 
The other variables $\alpha_{\rm s}$, $n_{\rm t}$ and $\alpha_{\rm t}$ 
are functions of these free parameters, so they are non-vanishing.
We also tried the case where the runnings $\alpha_{\rm s}$ and 
$\alpha_{\rm t}$ are set to 0 and found that the results 
are practically identical to those derived for $\epsilon_3=0$.

The consistency relations are different depending on the 
scenarios we study, see Eqs.\ (\ref{ratiob0}), (\ref{ratiob2}), 
(\ref{nt0}) and (\ref{nt2}). We run the CosmoMC code
for these four cases separately. 
As we will see, the likelihood results are 
insensitive to the change of the consistency relations.
The scalar power spectrum is constrained to be 
${\cal P}_{\rm s}(k_0) \approx 2.2 \times 10^{-9}$ 
for the pivot wave number (\ref{k0}).
This information can be used to place bounds on 
some parameters of the theories. 

We note that, in the context of slow-roll single-field 
inflation, the non-linear parameter $f_{\rm NL}^{\rm local}$ describing the 
scalar non-Gaussianities in the squeezed limit is of the order 
of the slow-roll parameters \cite{Malda,Cremi,TsuDe}. 
This is consistent with the recent Planck 
constraint $f_{\rm NL}^{\rm local}=2.7 \pm 5.8$ (68\,\% confidence-level, CL) \cite{Planck3}.
Since we focus on the slow-roll single-field framework, the local 
non-Gaussianities do not provide additional bounds on the 
models studied in this paper \cite{Cal8}.

References \cite{Smith,TsuLiddle,TSM,CT04} followed the same method 
but applied to one of the first datasets of precision cosmology, 
the first-year release of WMAP \cite{WMAP1}. 
In Refs.~\cite{Smith,TsuLiddle}, it was shown that, in the RS braneworld, the quadratic 
potential $V\propto \phi^2$ was inside the 95\,\%\,CL boundary but outside the 68\,\%\,CL contour. 
In the GB braneworld, the quadratic potential entered the 68\,\%\,CL region constrained 
by WMAP1 data \cite{TSM}. The same happened for the quartic potential $V\propto \phi^4$, but 
in the GB limit ($\delta/\delta_0 \gg 1$) the model was outside the 95\,\%\,CL boundary.
In Ref.~\cite{CT04}, two of the present authors studied observational constraints on 
hybrid scenarios of non-commutative braneworlds. The effect of non-commutativity allowed the possibility, with WMAP1 data, to rescue models 
disfavored in standard GR. 

Later studies on braneworld and non-commutative inflation exploited more recent 
data but without extracting the likelihood bounds from new numerical simulations. 
The RS braneworld was compared against WMAP3 \cite{MuM,BGS} and 
WMAP5 data \cite{BZCB}: the potentials with small-field variations turned out to be favored \cite{BGS} and the quartic potential excluded \cite{MuM}. The quartic potential was also excluded in the GB braneworld by WMAP3 data \cite{MuM}. Planck data were recently used also in \cite{NoR}, but on a DGP brane model different from the present ones. Reference \cite{MuM} also studied non-commutative inflation in the IR limit, using the patch formalism of \cite{Cal3} and WMAP3 data, which led to the exclusion of quadratic and quartic potentials in the $\sigma=6$ model. These potentials have been ruled out also by Planck data, but only in the UV mild non-commutative regime \cite{LiZ13}. All of these works based their conclusions (which agree with ours, whenever a comparison is possible) on the bounds on the scalar index, on its running and on the tensor-to-scalar ratio, sometimes reusing the likelihood contour plots of \cite{CT04} or of the WMAP and Planck teams. In this respect, the present 
 analysis constitutes the first significant update on braneworld and non-commutative scenarios in the post-WMAP era.

\section{Observational constraints on braneworld inflation} \label{braneconsec}

We study observational constraints on several inflaton 
potentials in the context of RS ($q=2$) and GB ($q=2/3$) 
braneworlds in commutative spacetime 
($\sigma=0$). Under the slow-roll approximation 
($\dot{\phi}^2/2 \ll V(\phi)$ and $|\ddot{\phi}| \ll |3H\dot{\phi}|$), 
Eqs.~(\ref{Hubble}) and (\ref{phieq}) read
\ba
& & H^2 \approx \beta_q^2 V^q\,,\label{slow1}\\
&& 3H \dot{\phi}+V_{,\phi} \approx 0\,.
\label{slow2}
\ea
Taking the time derivative of Eq.~(\ref{slow1}) and using Eq.~(\ref{slow2}), 
the slow-roll parameter $\epsilon_{\phi}=\epsilon_1=-\dot{H}/H^2$ reads
\be
\epsilon_{\phi}=\frac{qV_{,\phi}^2}{6\beta_q^2V^{q+1}}\,.
\label{slowpara1}
\ee
Similarly, we also obtain the following relation:
\be
\eta_{\phi}=-\epsilon_{\phi}
+\frac{V_{,\phi \phi}}{3\beta_q^2 V^q}\,.
\label{slowpara2}
\ee

The scalar power spectrum (\ref{Ps}) can be expressed as 
\be
{\cal P}_{\rm s}=\frac{9\beta_q^6 V^{3q}}{25\pi^2V_{,\phi}^2}\,.
\label{Ps2}
\ee
On using Eqs.~(\ref{ep1}) and (\ref{ep2}), the scalar spectral 
index (\ref{nsb}) and the tensor-to-scalar ratio (\ref{ratiob1}) read
\ba
n_{\rm s} &=& 1-\frac{1}{3\beta_q^2 V^q} \left(3q 
\frac{V_{,\phi}^2}{V} -2V_{,\phi \phi} \right)\,,
\label{nsbrane} \\
r &=& \frac{8q}{3\beta_q^2 \zeta_q} 
\frac{V_{,\phi}^2}{V^{q+1}}\,.
\label{rbrane}
\ea
Under the slow-roll approximation, the number of e-foldings from 
the end of inflation (field value $\phi_{\rm f}$) to the epoch with
the field value $\phi$ is given by 
\be
N=3\beta_q^2 \int_{\phi_{\rm f}}^{\phi} d\tilde{\phi}\,\frac{V^q}{V_{,\tilde{\phi}}}\,.
\label{efold}
\ee

In the following, we show observational constraints on the models for $N=50$ and $60$. Note that the value of $N$ can be
different depending on the cosmological history 
after inflation \cite{Liddle:2003as}. 
One possible effect on the value of $N$ comes
from the reheating stage after inflation. 
If reheating is caused by a perturbative decay of $\phi$, 
the Universe is matter dominated  
for the inflaton potential approximated as 
$V(\phi) \simeq m^2(\phi-\phi_0)^2/2$ around the minimum. 
In the framework of GR, the number of e-foldings is modified 
depending on the reheating temperature $T_{\rm reh}$, as
\be
N\simeq 56-\frac{2}{3}\ln\frac{10^{16}\,{\rm GeV}}
{\rho_{\rm end}^{1/4}}-\frac{1}{3}\ln\frac{10^9\,{\rm GeV}}{T_{\rm reh}}\,,
\label{efolds}
\ee
where $\rho_{\rm end}$ is the energy density of the Universe at the
end of inflation.  

In the RS braneworld scenario, the $\rho^2$ term in the background
equation could dominate during reheating and one may expect a change
in the value of $N$. If the $\rho^2$-dominated phase, where
$H^2\propto a^{-6}$, ends before the
  completion of reheating, it gives rise to a change $\Delta
N=(1/6)\ln(\rho_{\rm end}/\rho_T)$ \cite{Copeland:2005qe}, where
$\rho_T=2\lambda$ represents the energy density of the Universe at the
transition from $\rho^2$ to $\rho$ dominating regime.  Because of the
coefficient $1/6$ and the logarithmic dependence of the ratio
$\rho_{\rm end}/\rho_T$, the modification to $N$ cannot be so large.
In non-commutative models, the background equation is not modified and
reheating proceeds in a similar way as in GR.  Therefore, in
most cases $N$ ranges between 50 and 60. 
For completeness, below we will quote experimental bounds 
also for lower and higher values of $N$.

\subsection{Monomial inflation}

Let us first study monomial inflation \cite{chaotic} 
with the monomial potential 
\be
V(\phi)=V_0 \phi^p\,,
\label{monomialpo}
\ee
where $V_0$ and $p$ are positive constants. 
The field value at the end of inflation is determined by 
the condition $\epsilon_{\phi} (\phi_{\rm f})=1$, i.e., 
$\phi_{\rm f}^{p(q-1)+2}=p^2 qV_0^{1-q}/(6\beta_q^2)$.
Integrating Eq.~(\ref{efold}), the field $\phi$ can be 
expressed in terms of $N$, as
\be
\phi^{p(q-1)+2}=\frac{p}{6\beta_q^2 V_0^{q-1}}
\left[ 2(pq-p+2)N+pq \right]\,.
\ee
{}From Eqs.~(\ref{nsbrane}) and (\ref{rbrane}) we obtain 
\ba
n_{\rm s} &=& 1-\frac{p(6q-4)+4}{2N(pq-p+2)+pq}\,,
\label{nscha} \\
r &=& \frac{16pq}{\zeta_q}
\frac{1}{2N(pq-p+2)+pq}\,.
\label{rcha}
\ea

We discuss the RS and GB cases separately.

\subsubsection{RS braneworld}

In the RS case, the observables (\ref{nscha}) and 
(\ref{rcha}) reduce to 
\ba
n_{\rm s} &=& 1-\frac{2(2p+1)}{N(p+2)+p}\,,
\label{ns2} \\
r &=& \frac{24p}{N(p+2)+p}\,.
\label{r2}
\ea
When $N=60$, these are given by 
$n_{\rm s}=0.9669$, $r=0.1326$ for $p=1$, 
$n_{\rm s}=0.9587$, $r=0.1983$ for $p=2$, and 
$n_{\rm s}=0.9505$, $r=0.2637$ for $p=4$, respectively.

In Fig.~\ref{fig:case2}, we plot the observational contours in the
$(n_{\rm s}, r)$ plane derived by the joint data analysis of
Planck+WP+BAO+high-$\ell$. We also show the theoretical values
(\ref{ns2}) and (\ref{r2}) for $p=1,2,4$ with $N=50, 60$.  The
potentials $V(\phi)=V_0\phi^p$ with $p \geq 2$ are outside 
the 95\,\%\,CL observational contour for both $N=50$ and 60.
For $N>70$, the theoretical line goes outside the 95\,\%\,CL
region. Although a small $N$ may be allowed by modifying 
the reheating scenario, for $p=1,2,4$ the model is outside 
the 95\,\%\,CL boundary when $N<52$.

The linear potential $V(\phi)=V_0 \phi$ arises for a string axion in type IIB 
compactification in the presence of wrapped branes \cite{linear}. 
This case is marginally inside the 
95\,\%\,CL boundary for $N=60$, but it is outside the 95\,\%\,CL 
region for $N=50$.
These tensions come from the fact that the tensor-to-scalar 
ratio gets even larger than that in GR for the same value of $p$.

Note that the exponential potential 
$V(\phi)=V_0 e^{-\alpha \phi/m_{\rm Pl}}$ \cite{exponential}
corresponds to the limit $p \to \infty$, i.e., 
$n_{\rm s}=1-4/(N+1)$ and $r=24/(N+1)$.
In this case, the model is far outside the 95\,\%\,CL 
observational boundary.

\subsubsection{GB braneworld}

In the GB case,
\ba
n_{\rm s} &=& 1-\frac{6}{N(6-p)+p}\,,
\label{ns3} \\
r &=& \frac{16p}{N(6-p)+p}\,.
\label{r3}
\ea
When $N=60$, we have 
$n_{\rm s}=0.9801$, $r=0.0532$ for $p=1$, 
$n_{\rm s}=0.9752$, $r=0.1322$ for $p=2$, and 
$n_{\rm s}=0.9516$, $r=0.5161$ for $p=4$, respectively.

\begin{figure}[H]
  \begin{minipage}[t]{0.47\hsize}
    \begin{center}
    \includegraphics[height=2.5in,width=2.8in]{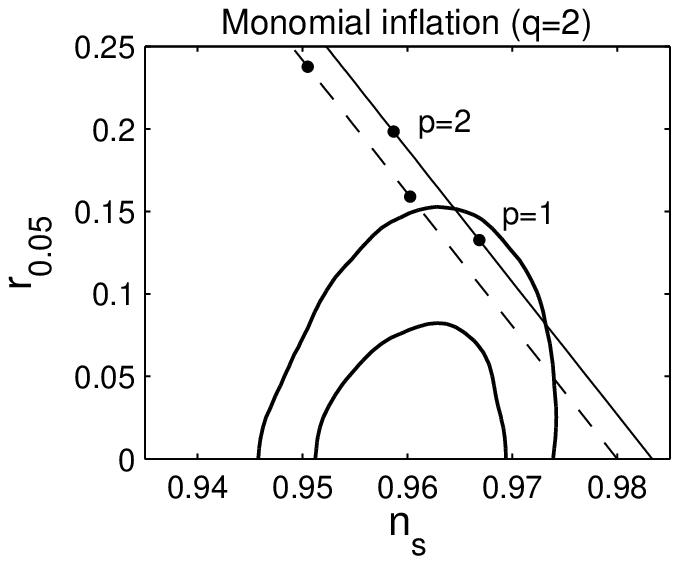}\\
       \includegraphics[height=2.5in,width=2.8in]{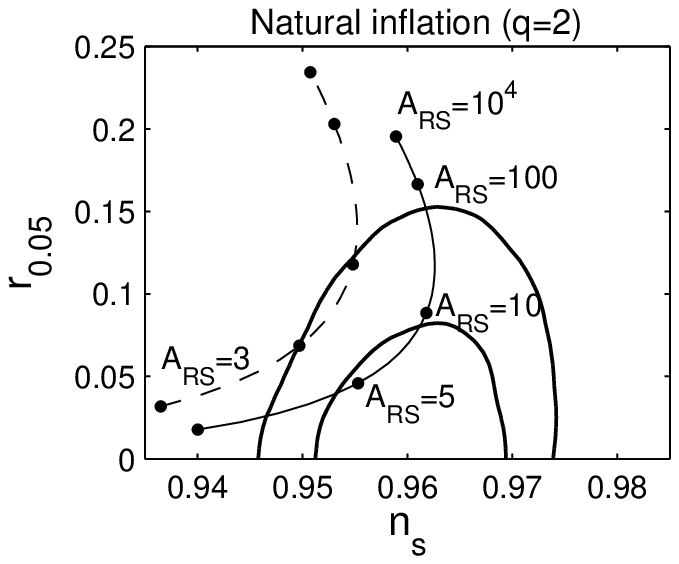}\\
        \includegraphics[height=2.5in,width=2.8in]{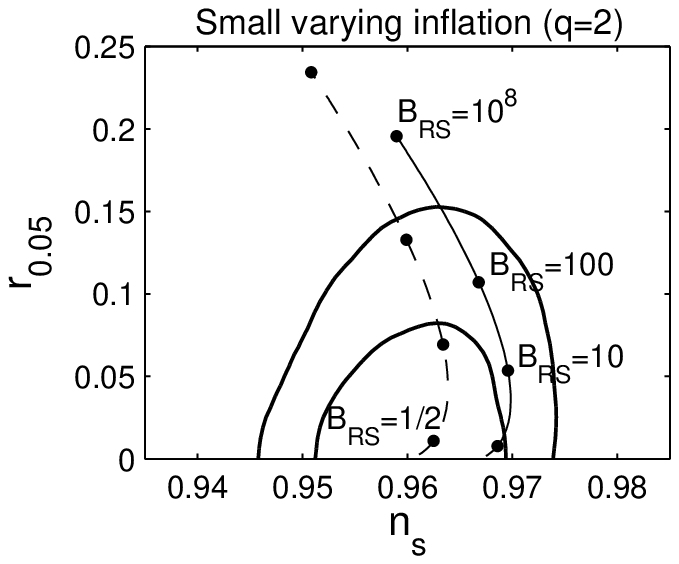}
      \caption{
Two-dimensional observational constraints on the RS braneworld ($q=2$)
in the ($n_{\rm s}, r$) plane.
Each panel corresponds to monomial inflation (top), 
natural inflation (middle), and small-varying inflaton models (bottom).
The consistency relation $r=-8n_{\rm t}$ is assumed. 
The two contours show the 68\,\% (inside) and 95\,\% (outside) 
CL boundaries, respectively. 
The solid and dashed curves correspond to the 
theoretical predictions  for
$N=60$ and 50, respectively.
        \label{fig:case2}
      }
    \end{center}
  \end{minipage}
\hfill
  \begin{minipage}[t]{0.48\hsize}
    \begin{center}
     \includegraphics[height=2.5in,width=2.8in]{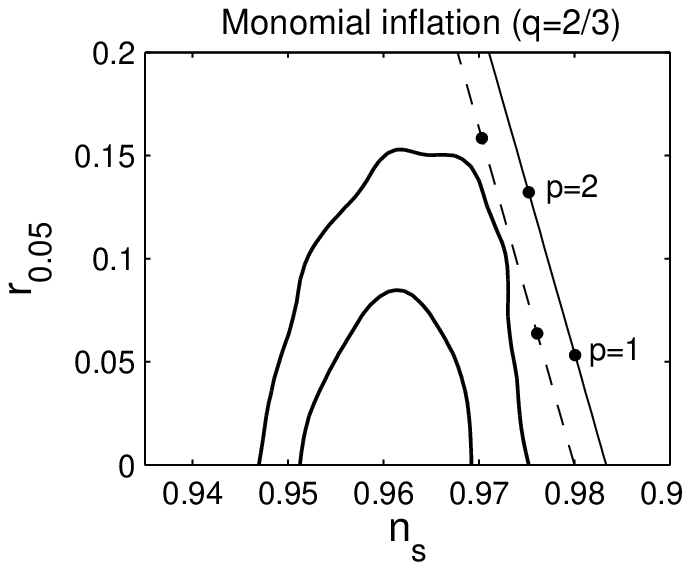}\\
       \includegraphics[height=2.5in,width=2.8in]{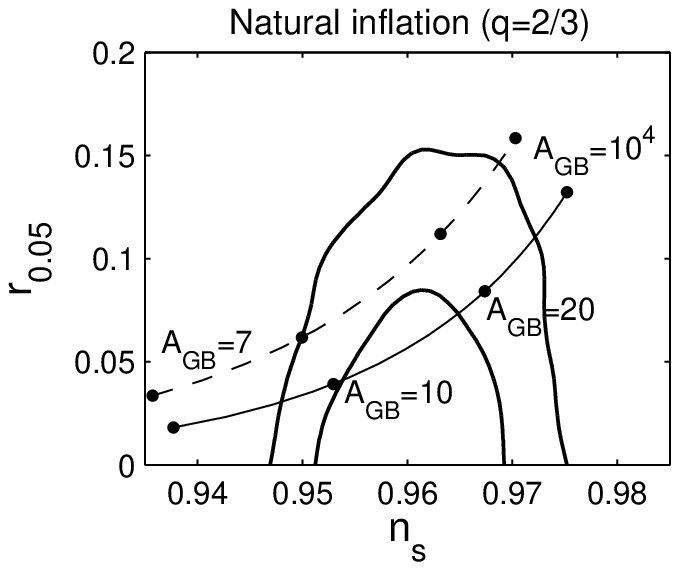}\\
        \includegraphics[height=2.5in,width=2.8in]{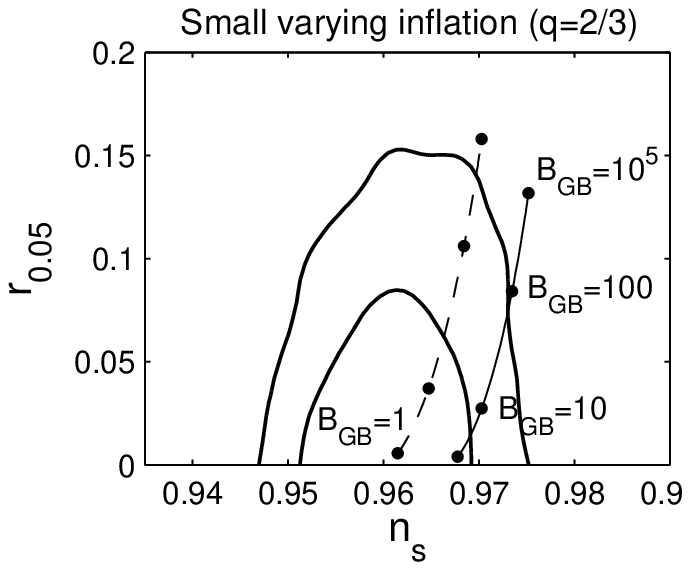}
         \caption{                  
Two-dimensional observational constraints on the GB braneworld 
($q=2/3$) in the ($n_{\rm s}, r$) plane for monomial inflation (top), natural 
inflation (middle), and small-varying inflaton models (bottom).
The consistency relation $r=-16n_{\rm t}$ is assumed. 
The meanings of observational contours and theoretical curves are 
the same as shown in Fig.~\ref{fig:case2}.
        \label{fig:case3}
      }
    \end{center}
  \end{minipage}
\end{figure}

As we see in Fig.~\ref{fig:case3}, the potentials with 
$p>0$ are outside the 95\,\%\,CL boundary.
Even for the linear potential ($p=1$) where $r$ is smaller than 0.1, 
the model is in tension with the data because of 
the tight upper bound on $n_{\rm s}$.
The model is excluded for $N>48$ at the 95\,\%\,CL.
Thus, monomial inflation is observationally disfavored  
in both the RS and the GB case.

\subsection{Natural inflation}

Natural inflation \cite{natural} is characterized 
by the potential 
\be
V(\phi)=V_0 \left[ 1+\cos (\phi/f) \right]\,,
\label{napotential}
\ee
where $V_0$ and $f$ are positive constants. 
In particular, $f$ is the energy scale at which the global symmetry 
associated with this model is broken.

\subsubsection{RS braneworld}

In the RS braneworld, 
the slow-roll parameter $\epsilon_{\phi}$ is given by 
\be
\epsilon_{\phi}=\frac{1}{3A_{\rm RS}} \frac{1-x}{(1+x)^2}\,,
\ee
where $A_{\rm RS}=\beta_2^2 V_0 f^2$ and $x:=\cos (\phi/f)$. 
Since we focus on the regime $0<\phi<\pi f$, the variable $x$ 
is in the range $-1<x<1$.
The coefficient $\beta_2^2$ in Eq.~(\ref{Hubble}) is related 
to the brane tension $\lambda$ and the 4-dimensional 
Planck mass $m_{\rm Pl}$, as 
$\beta_2^2=4\pi/(3m_{\rm Pl}^2 \lambda)$ \cite{Shiromizu}.
Then, the parameter $A_{\rm RS}$ can be expressed as
\be
A_{\rm RS}=\frac{4\pi}{3} \frac{V_0}{\lambda} 
\left( \frac{f}{m_{\rm Pl}} \right)^2.
\ee

We are considering the high-energy regime in which the 
$\rho^2/\lambda$ term dominates over $\rho$, i.e., 
$V_0/\lambda \gg 1$. In order to realize a sufficient amount of 
inflation, we require that $A_{\rm RS}>O(1)$ 
(unless $\phi$ be initially extremely close to 0). 
In GR, the symmetry-breaking scale $f$ is constrained to be 
$f>0.9m_{\rm Pl}$ from the Planck data \cite{Kuro}. 
In the RS braneworld, it is possible to realize inflation 
even when $f$ is smaller than $m_{\rm Pl}$.

The end of inflation corresponds to $x_{\rm f}=-1+(\sqrt{24A_{\rm RS}
+1}-1)/(6A_{\rm RS})$, 
where $x_{\rm f}=\cos(\phi_{\rm f}/f)$. 
The number of e-foldings is given by 
\be
N=6A_{\rm RS} \ln \frac{1-x_{\rm f}}{1-x}
+3A_{\rm RS} (x_{\rm f}-x)\,.
\label{efoldna}
\ee
From Eqs.~(\ref{nsbrane}) and (\ref{rbrane}), we have 
\ba
n_{\rm s} &=&1-\frac{2}{3A_{\rm RS}} \frac{3-2x}{(1+x)^2}\,,
\label{nsna} \\
r &=& \frac{8}{A_{\rm RS}} \frac{1-x}{(1+x)^2}\,.
\label{rna}
\ea
For a given $A_{\rm RS}$, we can numerically obtain the values of
$x$ at $N=50$ and 60 by using Eq.~(\ref{efoldna}).
Then, the scalar spectral index and the tensor-to-scalar 
ratio are evaluated from Eqs.~(\ref{nsna}) and (\ref{rna}).

In the limit that $A_{\rm RS} \gg 1$, we have $x_{\rm f} \to -1$ and 
$x \to 1+2W[-\rme^{-(N+6A_{\rm RS})/(6A_{\rm RS})}]$ from Eq.~(\ref{efoldna}), 
where $W(z)$ is the Lambert $W$ function. 
Substituting this into Eqs.~(\ref{nsna})-(\ref{rna}) and taking 
the limit $A_{\rm RS} \to \infty$, it follows that 
$n_{\rm s} \to 1-5/(2N)$ and $r \to 12/N$ 
($n_{\rm s}=0.9583$ and $r=0.2$ for $N=60$).
These correspond to the values (\ref{ns2}) and (\ref{r2}) 
with $p=2$ and $N \gg 1$, so that the model approaches
monomial inflation with the potential $V=V_0 \phi^2$ for 
$A_{\rm RS} \to \infty$. 

For decreasing $A_{\rm RS}$, the tensor-to-scalar ratio 
gets smaller. In the regime $A_{\rm RS} \gg 1$,
the scalar spectral index gets larger as 
$A_{\rm RS}$ decreases, but it starts to decrease 
for $A_{\rm RS} \lesssim 10$. 
In Fig.~\ref{fig:case2}, we plot the numerical values of $n_{\rm s}$ and $r$ 
as a function of $A_{\rm RS}$ in the range 
$3 \le A_{\rm RS} \le 10^4$ for $N=50$ and 60.
From the joint data analysis of Planck+WP+BAO+high-$\ell$,
the parameter $A_{\rm RS}$ is constrained to be 
\ba
&&
3.5<A_{\rm RS}<49 \quad (95\,\%\,{\rm CL}) \quad 
{\rm for} ~N=60,\label{Abound}\\
&&
5.0<A_{\rm RS}<11 \quad (95\,\%\,{\rm CL}) \quad 
{\rm for} ~N=50.
\ea
The constraints on $A_{\rm RS}$ become weaker at larger $N$.  
For example, $3.1 < A_{\rm RS} < 75 \, (95\,\%\,{\rm CL})$ for $N=80$. 
For small values of $N$, the model is excluded at 95$\%$ CL if $N<48$.  
The upper and lower bounds on $A_{\rm RS}$
come from the constraints on $r$ and $n_{\rm s}$, respectively.  The
bound (\ref{Abound}) can be recast as
$0.91<\sqrt{V_0/\lambda}\,(f/m_{\rm Pl})<3.4$, so a symmetry-breaking
scale $f$ smaller than $m_{\rm Pl}$ can be allowed for $V_0/\lambda
\gg 1$.  We note that, only when $N>57$, there are some parameter
values in which the model is inside the 68\,\%\,CL observational
contour.

\subsubsection{GB braneworld}

In the GB braneworld, the slow-roll parameter reads
\be
\epsilon_{\phi}=\frac{1}{9A_{\rm GB}} \frac{1-x}{(1+x)^{2/3}}\,,
\ee
where $A_{\rm GB}=\beta_{2/3}^2f^2V_0^{-1/3}$.
The number of e-foldings is given by 
\be
N=3A_{\rm GB} \int_{x_{\rm f}}^{x}\rmd\tilde{x}\,\frac{(1+\tilde{x})^{2/3}}
{1-\tilde{x}^2} \,,
\label{NGB}
\ee
where $x_{\rm f}=\cos(\phi_{\rm f}/f)$ is known by the 
condition $\epsilon_{\phi}(x_{\rm f})=1$.
Equation (\ref{NGB}) can be analytically integrated, 
but we do not write its explicit expression 
because of its complexity.

The scalar spectral index and the tensor-to-scalar ratio are
\ba
n_{\rm s} &=& 1-\frac{2}{3A_{\rm GB}} \frac{1}{(1+x)^{2/3}}\,,
\label{nsgb} \\
r &=& \frac{16}{9A_{\rm GB}} \frac{1-x}
{(1+x)^{2/3}}\,.
\label{rgb}
\ea
For a given $A_{\rm GB}$, the values of $x$ corresponding to 
$N=50$ and 60 are known from Eq.~(\ref{NGB}), so that 
$n_{\rm s}$ and $r$ are evaluated from Eqs.~(\ref{nsgb}) and (\ref{rgb}).

In Fig.~\ref{fig:case3}, we plot the theoretical values of $n_{\rm s}$ and
$r$ in the range $7 \le A_{\rm GB} \le 10^4$. 
When $A_{\rm GB}=7$ and $N=60$, we have that 
$n_{\rm s}=0.9377$ and $r=0.0181$, 
in which case the model is outside the 95\,\%\,CL boundary.
For increasing $A_{\rm GB}$, both $n_{\rm s}$ and $r$ get larger, 
so that the model enters the 95\,\%\,CL region. 
In the limit that $A_{\rm GB} \to \infty$, $n_{\rm s}$ and $r$ 
approach the values (\ref{ns3}) and (\ref{r3})
with $p=2$.  In this limit, the model is again outside the 95\,\%\,CL
contour. Then, the parameter $A_{\rm GB}$ is constrained to be
\ba
&&8.7<A_{\rm GB}<38 \quad (95\,\%\,{\rm CL}) \quad {\rm for} ~N=60,\\
&&10<A_{\rm GB}<55 \quad~(95\,\%\,{\rm CL}) \quad {\rm for} ~N=50.
\label{AboundGB}
\ea
For larger $N$, the constraints on $A_{\rm GB}$ become 
stronger. For example, we obtain $8.1 < A_{\rm GB} < 22 \, (95\,\%\,{\rm CL})$ 
for $N=80$. With small values of $N$, the model is excluded 
at 95$\%$\,CL if $N<42$. If $N>53$, then there is some
non-trivial parameter space in which the model is inside 
the 68\,\%\,CL boundary.

\subsection{Small varying inflaton models}

In GR there are some inflationary models in which the variation 
of the field from the epoch at which the perturbations relevant 
to the CMB crossed the Hubble radius to the end of inflation
is smaller than the order of the Planck mass $m_{\rm Pl}$.
We call such models small-varying inflaton models.
We call that the situation of this small variation 
can be subject to change in the RS and GB 
braneworld scenarios.
For concreteness, we consider the following potential:
\be
V(\phi)=V_0 (1-e^{-\alpha \phi/m_{\rm Pl}} )^2\,,
\label{smallpo}
\ee
where $V_0$ and $\alpha$ are positive constants.
In the Starobinsky model, where the Lagrangian is given by
$f(R)=R+R^2/(6M^2)$ \cite{Star}, the potential 
in the Einstein frame corresponds to (\ref{smallpo}) with 
$V_0=3M^2m_{\rm Pl}^2/(32\pi)$ 
and $\alpha=4\sqrt{\pi/3}$ \cite{Maeda}.
In the following, when we mention the Starobinsky model, 
it means the potential $V(\phi)=V_0 (1-e^{-4\sqrt{\pi/3}\,\phi/m_{\rm Pl}} )^2$ 
in the Einstein frame.

\subsubsection{RS braneworld}

In the RS case, the slow-roll parameter is given by 
\be
\epsilon_{\phi}=\frac{4}{3B_{\rm RS}} \frac{y^2}{(1-y)^4}\,,
\ee
where $B_{\rm RS}=\beta_2^2V_0 m_{\rm Pl}^2/\alpha^2$ and 
$y=e^{-\alpha \phi/m_{\rm Pl}}$. 
The parameter $B_{\rm RS}$ can be expressed in terms of 
the brane tension $\lambda$, as 
\be
B_{\rm RS}=\frac{4\pi}{3} \frac{V_0}{\lambda}
\frac{1}{\alpha^2}\,.
\ee
In the Starobinsky model, it follows that $B_{\rm RS}=V_0/(4\lambda) \gg 1$. 
In general, inflation can occur even for $B_{\rm RS}=O(1)$ in the regime $y \ll 1$.

The field value $y_{\rm f}=e^{-\alpha \phi_{\rm f}/m_{\rm Pl}}$ at the end 
of inflation is known by numerically solving 
$\epsilon_{\phi}(y_{\rm f})=1$ for a given $B_{\rm RS}$. 
The number of e-foldings is 
\be
N=\frac34 B_{\rm RS}
\left[ (y_{\rm f}-y)\left(6-y_{\rm f}-y+\frac{2}{yy_{\rm f}} \right)+
6 \ln \frac{y}{y_{\rm f}} \right]\,,
\ee
by which the value of $y$ can be found at $N=50$ and 60.
The scalar spectral index and the tensor-to-scalar ratio are
\ba
n_{\rm s} &=&1-\frac{4}{3B_{\rm RS}} \frac{y(4y+1)}{(1-y)^4}\,,
\label{nssta} \\
r &=& \frac{32}{B_{\rm RS}} \frac{y^2}{(1-y)^4}\,.
\label{rsta}
\ea
For increasing $B_{\rm RS}$, the tensor-to-scalar ratio gets larger, 
whereas $n_{\rm s}$ does not change significantly.
When $N=60$, for example, we obtain
$n_{\rm s}=0.9692$, $r=0.0136$ for $B_{\rm RS}=1$,
$n_{\rm s}=0.9696$, $r=0.0536$ 
 for $B_{\rm RS}=10$, 
and $n_{\rm s}=0.9668$, $r=0.1071$ for $B_{\rm RS}=100$.

In the limit $B_{\rm RS} \to \infty$, $n_{\rm s}$ and $r$ approach the 
values of monomial inflation with the potential 
$V(\phi)=V_0 \phi^2$ (see Fig.~\ref{fig:case2}).
This is due to the fact that, for larger $B_{\rm RS}$, inflation can be 
realized in the regime around the potential minimum at $\phi=0$.
This situation is analogous to what happens in natural inflation 
in the limit $A_{\rm RS} \to \infty$.
{}From the joint data analysis of Planck+WP+BAO+high-$\ell$
the parameter $B_{\rm RS}$ is constrained to be 
\ba
& &B_{\rm RS}<1650 \quad (95\,\%\,{\rm CL}) 
\quad {\rm for} ~N=60, \label{BRS} \\
& &B_{\rm RS}<170 \quad~\,(95\,\%\,{\rm CL}) 
\quad {\rm for} ~N=50.
\label{Bbound}
\ea
For large values of $N$, the model is excluded at 95$\%$ CL if $N>76$. 
For small $N$, we obtain $B_{\rm RS}<16 \,(95\,\%\,{\rm CL}) $ when $N=40$. 
The upper bound (\ref{BRS}) puts a constraint
on the ratio $V_0/\lambda$. In the Starobinsky model, for example, it
follows that $V_0/\lambda<6.6 \times 10^3$. Experiments have verified
GR down to scales $\lesssim 1\,\text{mm}$, corresponding to
$\lambda\gtrsim 10^{12}\,\text{GeV}^4$. Therefore, $V_0<7\times
10^{15}\,\text{GeV}^4$.

\subsubsection{GB braneworld}

In the GB case, the slow-roll parameter for the potential 
(\ref{smallpo}) is 
\be
\epsilon_{\phi}=\frac{4}{9B_{\rm GB}} \frac{y^2}
{(1-y)^{4/3}}\,,
\ee
where $B_{\rm GB}=\beta_{2/3}^2 m_{\rm Pl}^2/(\alpha^2 V_0^{1/3})$. 
The number of e-foldings is given by 
\be
N=\frac{3B_{\rm GB}}{2} \int_{y}^{y_{\rm f}} \rmd\tilde{y}\,
\frac{(1-\tilde{y})^{1/3}}{\tilde{y}^2}\,,
\ee
where $y_{\rm f}=e^{-\alpha \phi_{\rm f}/m_{\rm Pl}}$ is determined by the 
condition $\epsilon_{\phi}(y_{\rm f})=1$.
The observables are 
\ba
n_{\rm s} &=&1-\frac{4}{3B_{\rm GB}} \frac{y}{(1-y)^{4/3}}\,,
\label{nsstagb} \\
r &=& \frac{64}{9B_{\rm GB}} \frac{y^2}{(1-y)^{4/3}}\,.
\label{rstagb}
\ea

When $B_{\rm GB}=1$ and $N=60$, we have  $n_{\rm s}=0.9678$ 
and $r=0.0040$, 
 so the model is well inside 
the 68\,\%\,CL region.
For larger $B_{\rm GB}$, the tensor-to-scalar ratio gets larger, 
whereas $n_{\rm s}$ increases a bit (see Fig.~\ref{fig:case3}).
In the limit $B_{\rm GB} \to \infty$, the observables (\ref{nsstagb}) 
and (\ref{rstagb}) approach those given 
in Eq.~(\ref{ns3}) and (\ref{r3}) with $p=2$, 
in which case the model is outside the 95\,\%\,CL 
boundary. Then, the parameter $B_{\rm GB}$ is 
constrained to be
\ba
&& B_{\rm GB}<67 \quad~\,(95\,\%\,{\rm CL}) \quad {\rm for} ~N=60,\\
& &B_{\rm GB}<580 \quad (95\,\%\,{\rm CL}) \quad {\rm for} ~N=50.
\label{BboundGB}
\ea
For large values of $N$, the model is excluded at 95$\%$ CL if $N>80$. 
For small $N$, we get $B_{\rm GB}< 160 \,(95\,\%\,{\rm CL}) $ for $N=40$.

\section{Observational constraints on non-commutative inflation} 
\label{branenoncomsec}

We proceed to observational constraints on non-commutative 
inflation ($\sigma=2$ and 6) for $q=1$. 
We can employ the same slow-roll equations of motion as 
(\ref{slow1}) and (\ref{slow2}), so that the slow-roll parameters are 
given by (\ref{slowpara1}) and (\ref{slowpara2}).
The scalar power spectrum is
\be
{\cal P}_{\rm s}=\frac{9\beta_1^6 V^3}{25\pi^2V_{,\phi}^2} 
\Sigma^2\,,
\ee
where $\beta_1^2=8\pi/(3m_{\rm Pl}^2)$. 
The scalar spectral index and the 
tensor-to-scalar ratio are
\ba
n_{\rm s} &=& 1-\frac{1}{3\beta_1^2 V} \left(
\frac{6-\sigma}{2} \frac{V_{,\phi}^2}{V}
-2V_{,\phi \phi} \right),\label{nsnon} \\
r &=& \frac{8}{3\beta_1^2} \frac{V_{,\phi}^2}{V^2}\,.
\label{rnon}
\ea
The number of e-foldings is given by 
$N=3\beta_1^2 \int_{\phi_{\rm f}}^{\phi} \rmd\tilde{\phi}\,
V/V_{,\tilde{\phi}}$.

When $\sigma=6$, we have $n_{\rm s}=1+2V_{,\phi \phi}/(3\beta_1^2 V)$, 
so that the spectrum is blue-tilted ($n_{\rm s}>1$) for potentials with 
positive curvature ($V_{,\phi \phi}>0$). 
This is the case of monomial inflation.
For natural inflation and small-varying inflaton models, 
there exist some field ranges with negative curvature.
In the following, we study the same three inflaton 
potentials discussed in Sec.~\ref{braneconsec}.

\subsection{Monomial inflation}

For the potential (\ref{monomialpo}), the field value at the end of 
inflation is given by $\phi_{\rm f}=p/(\sqrt{6}\beta_1)$. 
Since $\phi$ is related to $N$ via $\beta_1^2 \phi^2
=2p (N+p/4)/3$, the observables (\ref{nsnon}) 
and (\ref{rnon}) are
\ba
n_{\rm s} &=& 1+\frac{p(\sigma-2)-4}{4N+p}\,,
\label{nsnoncha} \\
r &=& \frac{16p}{4N+p}\,,
\label{rnoncha}
\ea
where $r$ is the same as in standard GR.

If $\sigma=6$, then $n_{\rm s}=1+4(p-1)/(4N+p)$, 
so that the spectrum is blue-tiled for $p>1$. 
When $N=60$, we have $n_{\rm s}=1.0165$, $r=0.1322$ 
for $p=2$ and $n_{\rm s}=1.0492$, $r=0.2623$ for $p=4$. 
If $\sigma=2$, then $n_{\rm s}=1-4/(4N+p)<1$. 

In Figs.~\ref{fig:case4} and \ref{fig:case5}, we plot the theoretical
values of $n_{\rm s}$ and $r$ for $p=1,2,4$ and $\sigma=2,6$ with two
different values of $N$. Non-commutative monomial inflation with
$\sigma=6$ is outside the 95\,\%\,CL region because of the blue-tilted
spectrum. Interestingly, even the potentials with $\sigma=2$ are
outside the 95\,\%\,CL boundary. This comes from the fact that,
independent of the power $p$, the scalar spectral index is 
$n_{\rm s} \approx 1-1/N>0.98$ for $N>50$.
The theoretical lines shown in Figs.~\ref{fig:case4} and \ref{fig:case5} 
for $p>0$ are actually outside the 99\,\%\,CL boundary both 
for $\sigma=6$ and $2$.
For smaller $N$ the lines get closer to the allowed region.
However, they are outside the 95\,\%\,CL boundary 
even for $N=40$ in both cases $\sigma=6$ and $2$.

\subsection{Natural inflation}

In natural inflation described by the potential (\ref{napotential}), 
the field value at the end of inflation is given by 
$x_{\rm f}=\cos (\phi_{\rm f}/f)=(1-6A)/(1+6A)$, where 
$A=f^2 \beta_1^2=(8\pi/3)(f/m_{\rm Pl})^2$.
The field is related to the number of e-foldings, as
\be
x=1-\frac{12A}{1+6A}e^{-N/(3A)}\,.
\label{xna}
\ee
From Eqs.~(\ref{nsnoncha}) and (\ref{rnoncha}),
it follows that 
\ba
n_{\rm s} &=& 1-\frac{1}{6A}\frac{(\sigma-2)x+6-\sigma}{1+x}\,,
\label{nsnanon} \\
r &=& \frac{8}{3A} \frac{1-x}{1+x}\,.
\label{nanon}
\ea

Substituting Eq.~(\ref{xna}) into Eqs.~(\ref{nsnanon})-(\ref{nanon}) 
and taking the limit $A \to \infty$, we obtain 
$n_{\rm s} \to 1+(\sigma-4)/(2N+1)$ and $r \to 16/(2N+1)$.
These are equivalent to the values (\ref{nsnoncha}) 
and (\ref{rnoncha}) with $p=2$.
Then, in the limit $A \gg 1$, non-commutative natural inflation 
with $\sigma=2$ and 6 is in tension with observations 
because of the large scalar spectral index.

\begin{figure}[H]
  \begin{minipage}[t]{0.48\hsize}
    \begin{center}
     \includegraphics[height=2.5in,width=2.8in]{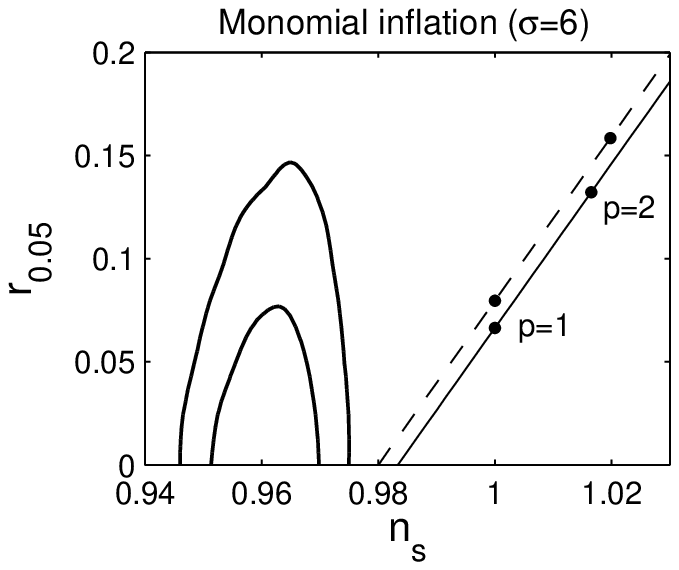}\\
       \includegraphics[height=2.5in,width=2.8in]{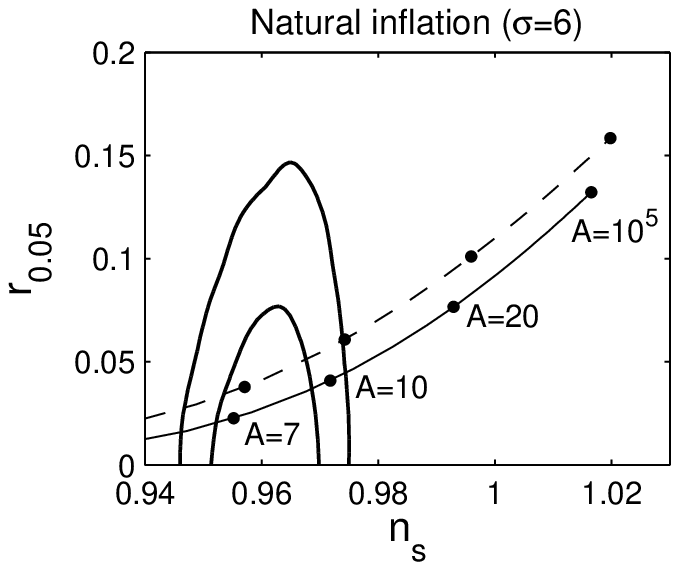}\\
        \includegraphics[height=2.5in,width=2.8in]{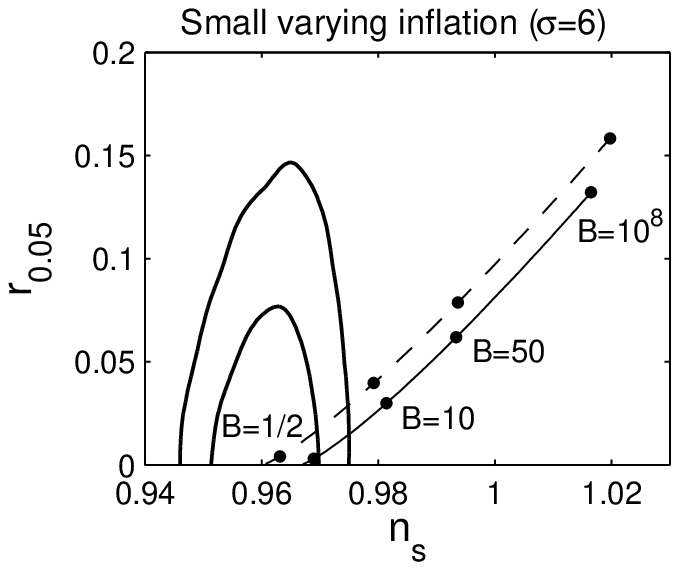}
      \caption{
Two-dimensional observational constraints on non-commutative inflation with 
$\sigma=6$ in the ($n_{\rm s},r$) plane. 
Each panel corresponds to monomial inflation (top), 
natural inflation (middle), and small-varying inflaton models (bottom).
The consistency relation $r=4n_{\rm t}$ is assumed. 
The two contours show the 68\,\% and 95\,\% CL boundaries. 
The solid and dashes curves represent the theoretical 
predictions of $n_{\rm s}$ and $r$ for $N=60$ and 50, 
respectively.
\label{fig:case4}
      }
    \end{center}
  \end{minipage}
\hfill
  \begin{minipage}[t]{0.48\hsize}
    \begin{center}
     \includegraphics[height=2.5in,width=2.8in]{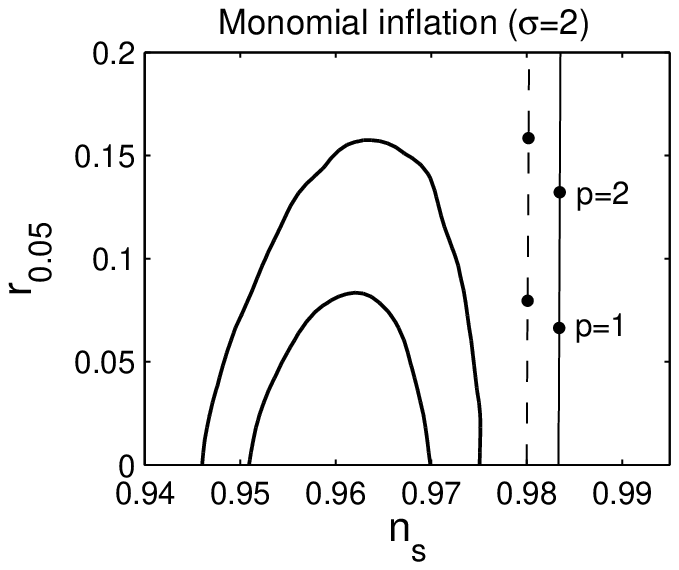}\\
       \includegraphics[height=2.5in,width=2.8in]{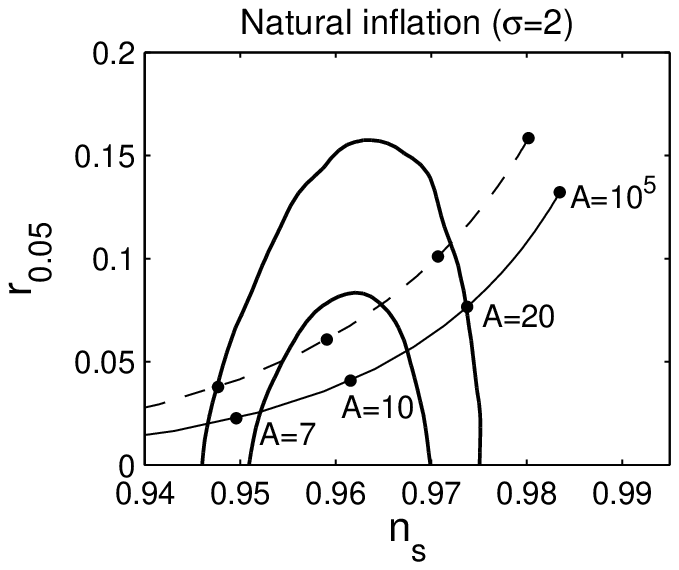}\\
        \includegraphics[height=2.5in,width=2.8in]{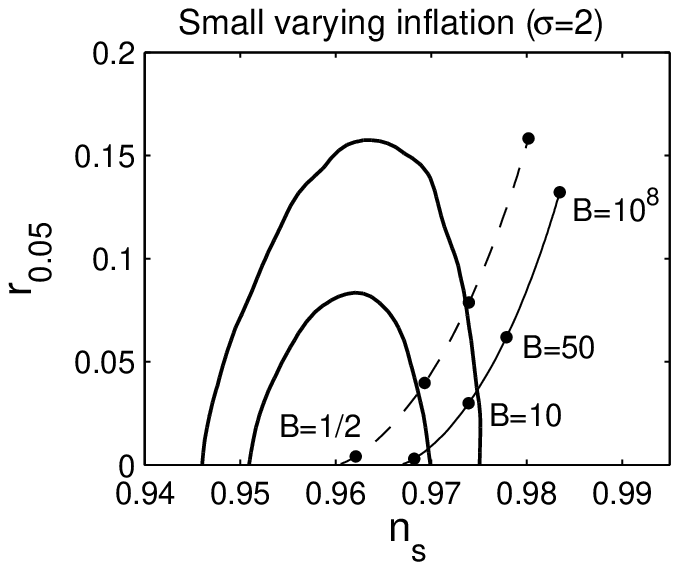}
      \caption{
Two-dimensional observational constraints on non-commutative inflation 
with $\sigma=2$ in the ($n_{\rm s},r$) plane for monomial inflation (top), 
natural inflation (middle), and small-varying inflaton models (bottom). 
The meanings of observational contours and theoretical curves are 
the same as shown in Fig.~\ref{fig:case4}.
      \label{fig:case5}
      }
    \end{center}
  \end{minipage}
\end{figure}

For decreasing $A$, both $n_{\rm s}$ and $r$ get smaller.
When $\sigma=6$ and $N=60$, for example, we have 
$n_{\rm s}=0.9856$ and $r=0.0627$ for $A=15$, 
$n_{\rm s}=0.9345$ and $r=0.0096$ for $A=5$. 
As we see in Figs.~\ref{fig:case4} and \ref{fig:case5}, 
there are some intermediate values 
of $A$ which are inside the 68\,\%\,CL observational contour.
From the joint data analysis of Planck+WP+BAO+high-$\ell$, 
the parameter $A$ is constrained to be 
\ba
\hspace{-0.8cm}
& &5.9<A<11 \quad (95\,\%\,{\rm CL}) \quad {\rm for} ~N=60 ~{\rm and} 
~\sigma=6\,,\label{Abo1} \\
\hspace{-0.8cm}
& &5.8<A<9.9 \quad (95\,\%\,{\rm CL}) \quad {\rm for} ~N=50 ~{\rm and} ~\sigma=6\,,
\ea
and
\ba
\hspace{-0.8cm}
&& 6.5<A<20 \quad (95\,\%\,{\rm CL}) \quad {\rm for} ~N=60 ~{\rm and} ~\sigma=2\,,
\label{Abo2}\\
\hspace{-0.8cm}
&& 7.0<A<23 \quad (95\,\%\,{\rm CL}) \quad {\rm for} ~N=50 ~{\rm and} ~\sigma=2\,.
\ea
When $N=80$, the constraints are $6.1<A<12 \, (95\,\%\,{\rm CL})$
for $\sigma=6$ and $6.2<A<17 \, (95\,\%\,{\rm CL})$ for $\sigma=2$.
When $N=40$, we get $5.7<A<8.9 \, (95\,\%\,{\rm CL})$ for $\sigma=6$
and $8.5<A<25 \, (95\,\%\,{\rm CL})$ for $\sigma=2$. 
The bounds (\ref{Abo1}) and (\ref{Abo2}) translate to $0.84<f/m_{\rm Pl}<1.15$
and $0.88<f/m_{\rm Pl}<1.55$, respectively, whose parameter ranges are
quite restrictive.  Moreover, the symmetry-breaking scale $f$ cannot
be much smaller than $m_{\rm Pl}$ as in the case of GR.

\subsection{Small varying inflaton models}

Finally, we study observational constraints on the potential (\ref{smallpo}). 
The end of inflation is characterized by $y_{\rm f}=(3B-\sqrt{6B})/(3B-2)$, 
where $y_{\rm f}=e^{-\alpha \phi_{\rm f}/m_{\rm Pl}}$ and 
$B=\beta_1^2m_{\rm Pl}^2/\alpha^2=8\pi/(3\alpha^2)$. 
The Starobinsky model in the Einstein frame corresponds 
to $\alpha=4\sqrt{\pi/3}$, i.e., $B=1/2$. 
The number of e-foldings is
\be
N=\frac{3B}{2} \left( \frac{1}{y}-\frac{1}{y_{\rm f}}
+\ln \frac{y}{y_{\rm f}}\right),
\label{smallN}
\ee
whereas $n_{\rm s}$ and $r$ are
\ba
n_{\rm s} &=& 1-\frac{2}{3B} \frac{y(2y-\sigma y+2)}{(1-y)^2}\,,
\label{nssm} \\
r &=& \frac{32}{3B} \frac{y^2}{(1-y)^2}\,.
\label{rsm} 
\ea

Expressing $y$ in terms of $N$ from Eq.~(\ref{smallN}), substituting 
it into Eqs.~(\ref{nssm}) and (\ref{rsm}), and taking the limit 
$B \to \infty$, we obtain the values (\ref{nsnoncha}) 
and (\ref{rnoncha}) with $p=2$. 
Therefore, this limit corresponds to the case of the quadratic potential, 
which lies outside the 95\,\%\,CL boundary for both $\sigma=2$ and 6.
As can be seen in Figs.~\ref{fig:case4} and \ref{fig:case5}, 
the models enter the observationally allowed 
region for smaller $B$. If $\sigma=6$ and $N=60$, then 
$n_{\rm s}=0.9934$ and $r=0.0620$ 
 for $B=50$, 
$n_{\rm s}=0.9814$ and $r=0.0299$ 
 for $B=10$, and 
$n_{\rm s}=0.9689$ and $r=0.0030$ 
 for $B=1/2$. 
The parameter $B$ is constrained to be 
\ba
&&B<3.5 \quad (95\,\%\,{\rm CL}) \quad {\rm for} ~N=60 ~{\rm and} ~\sigma=6\,,
\label{Bbou1} \\
&&B<5.9 \quad (95\,\%\,{\rm CL}) \quad {\rm for} ~N=50 ~{\rm and} ~\sigma=6\,,
\ea
and
\ba
&&B<14 \quad (95\,\%\,{\rm CL}) \quad {\rm for} ~N=60 ~{\rm and} ~\sigma=2\,,
\label{Bbou2}\\
&&B<45 \quad (95\,\%\,{\rm CL}) \quad {\rm for} ~N=50 ~{\rm and} ~\sigma=2\,.
\ea
The model is excluded if $N>80$ both for $\sigma=6$ and
$\sigma=2$. When $N=40$, we get $B<8.1 \, (95\,\%\,{\rm CL})$ for
$\sigma=6$ and $B<130 \, (95\,\%\,{\rm CL})$ for $\sigma=2$.

The Starobinsky model in the Einstein frame ($B=1/2$) is inside the 
68\,\%\,CL contour both for $\sigma=2$ and 6. 
The bounds (\ref{Bbou1}) and (\ref{Bbou2}) translate to 
$\alpha>1.55$ and $\alpha>0.77$, respectively. 
The models with $\alpha$ smaller than the order of 1 are 
disfavored because the cosmic acceleration relevant to 
CMB anisotropies occurs around the potential minimum.

\section{Conclusions} 
\label{concludesec}

We studied observational constraints on braneworld and non-commutative 
inflation in the light of the recent Planck data. 
The consistency relations between the tensor-to-scalar ratio $r$ and 
the tensor spectral index $n_{\rm t}$ are different depending on the scenario (RS braneworld, 
GB braneworld and two versions of non-commutative inflation). We ran the CosmoMC code for four different consistency relations
and found that the likelihood results are similar to those obtained 
in GR ($r=-8n_{\rm t}$). We also confirmed that, under the slow-roll approximation, 
the scalar and tensor runnings can be set to 0 in the likelihood analysis.

For each class of braneworld and non-commutative inflation, 
we placed experimental constraints on a number of representative 
inflaton potentials such as 
(i) monomial inflation: $V(\phi)=V_0 \phi^p$, 
(ii) natural inflation: $V(\phi)=V_0 [1+\cos(\phi/f)]$, and 
(iii) small-varying inflaton models: $V(\phi)=V_0 (1-e^{-\alpha \phi/m_{\rm Pl}})^2$.

In the RS braneworld, the monomial potential $V(\phi)=V_0 \phi^p$ 
is outside the 95\,\%\,CL boundary for $p \geq 2$.
The linear potential is marginally inside the 95\,\%\,CL border 
for $N=60$. The parameter $A_{\rm RS}=(4\pi/3)(V_0/\lambda)(f/m_{\rm Pl})^2$ 
in natural inflation is constrained to be $3.5<A_{\rm RS}<49$ (95\,\%\,CL) 
for $N=60$, so that a symmetry-breaking scale $f$ smaller 
than $m_{\rm Pl}$ can be allowed for $V_0/\lambda \gg 1$.
We note, however, that the allowed parameter space 
is quite narrow for $N=50$, i.e., $5.0<A_{\rm RS}<11$ (95\,\%\,CL). 
In small-varying inflaton models, the observables are quantified 
by the parameter $B_{\rm RS}=(4\pi/3)(V_0/\lambda)(1/\alpha^2)$.
Since $n_{\rm s}$ and $r$ approach the values of the quadratic potential 
$V(\phi)=V_0 \phi^2$ in the limit $B_{\rm RS} \to \infty$, the parameter
$B_{\rm RS}$ is constrained to be $B_{\rm RS}<1650$ (95\,\%\,CL). 
In the Starobinsky model ($\alpha=4\sqrt{\pi/3}$), this translates 
into the condition $V_0/\lambda<6.6 \times 10^{3}$.

In the GB braneworld, the monomial potential $V(\phi)=V_0 \phi^p$ 
with $p>0$ lies outside the 95\,\%\,CL region for $N \geq 50$. 
In natural inflation, the parameter $A_{\rm GB}=\beta_{2/3}^2f^2 
V_0^{-1/3}$ is constrained to be $8.7<A_{\rm GB}<38$ (95\,\%\,CL) 
for $N=60$, whereas in small-varying inflaton models the bound on 
the parameter $B_{\rm GB}=\beta_{2/3}^2m_{\rm Pl}^2/(\alpha^2 V_0^{1/3})$ 
is found to be $B_{\rm GB}<67$ (95\,\%\,CL) for $N=60$.

In non-commutative inflation with $\sigma=6$ and 2, 
the monomial potential $V(\phi)=V_0 \phi^p$ ($p>0$) is 
outside the 99\,\%\,CL boundary because the scalar 
spectral index gets larger than in GR. 
In natural inflation, the parameter $A=(8\pi/3)(f/m_{\rm Pl})^2$ 
is constrained to be $5.9<A<11$ (95\,\%\,CL) for 
$\sigma=6$, $N=60$ and $6.5<A<20$ (95\,\%\,CL) for 
$\sigma=2$, $N=60$. Hence the symmetry-breaking 
scale $f$ is of the order of $m_{\rm Pl}$ as in the case of GR.
In small-varying inflaton models, the bound on the parameter 
$B=8\pi/(3\alpha^2)$ is given by $B<3.5$ (95\,\%\,CL) 
for $\sigma=6$, $N=60$  and 
$B<14$ (95\,\%\,CL) for $\sigma=2$, $N=60$, so that the 
Starobinsky model in the Einstein frame ($B=1/2$) 
is consistent with the data. 

All these results have been also extended to values of $N$ smaller and larger 
than 50 and 60, depending on the details of the reheating stage. 
The corresponding bounds on the parameters of the inflationary potential 
(quoted in the text) are numerically different than those for $N=50,\,60$, 
but not enough to issue a qualitatively different physics.

Overall, braneworld and non-commutative models are not 
particularly favored over standard inflationary scenarios
by CMB experiments. 
The viable parameter space in those models is not large 
enough to give any significant advantage with respect to 
their GR counterpart. 
We note that there are some models which give rise to $n_{\rm s}$ 
smaller than 0.94 ---such as the minimal super-symmetric 
model \cite{MSSM}, renormalizable-inflection-point inflation \cite{infection}, 
tip inflation \cite{tip}, and so on. 
There may be some possibilities that models with small $n_{\rm s}$ 
were rescued by braneworld and non-commutative effects 
due to the increase of $n_{\rm s}$. 
We leave the analysis of such specific inflaton potentials 
for future work, possibly after the 2-year data release of Planck.

\begin{acknowledgments}
G.C., S.K.\ and S.T.\ acknowledge the i-Link cooperation program of CSIC (project ID i-Link0484) for partial sponsorship. The work of G.C.\ is under a Ram\'on y Cajal contract.
J.O.\ and S.T.\ are supported by the Scientific Research Fund of the JSPS (Nos.~23\,$\cdot$\,6781 and 24540286). S.T.\ also thanks for financial support the Scientific Research 
on Innovative Areas (No.~21111006). S.K.\ is supported by the Grant-in-Aid for Scientific research No.~24740149.
\end{acknowledgments}


\begin{thebibliography}{99}

%
\bibitem{WMAP1} 
D.N.\ Spergel {\it et al.}  [WMAP Collaboration],
\tia{First year Wilkinson Microwave Anisotropy Probe (WMAP) observations: Determination of cosmological parameters} \doinn{10.1086/377226}
{Astrophys.\ J.\ Suppl.}{148}{175}{2003} [\oarX{astro-ph/0302209}].
\bibitem{WMAP9} C.L.\ Bennett {\it et al.} [WMAP Collaboration], \tia{Nine-year Wilkinson Microwave Anisotropy Probe (WMAP) observations: final maps and results} \doinn{10.1088/0067-0049/208/2/20}{Astrophys.\ J.\ Suppl.}{208}{20}{2013} [\arX{1212.5225}];\\
G.\ Hinshaw {\it et al.} [WMAP Collaboration], \tia{Nine-year Wilkinson Microwave Anisotropy Probe (WMAP) observations: cosmological parameter results} \doinn{10.1088/0067-0049/208/2/19}{Astrophys.\ J.\ Suppl.}{208}{19}{2013} [\arX{1212.5226}].
\bibitem{Planck1} P.A.R.\ Ade {\it et al.} [Planck Collaboration], \tia{Planck 2013 results. I. Overview of products and scientific results} \arX{1303.5062}.
\bibitem{Planck2} P.A.R.\ Ade {\it et al.} [Planck Collaboration], \tia{Planck 2013 results. XXII. Constraints on inflation} \arX{1303.5082}.
\bibitem{Martin}
J.\ Martin, C.\ Ringeval and V.\ Vennin, 
\tia{Encyclopaedia Inflationaris} \arX{1303.3787}.
\bibitem{Ma}
Y.Z.\ Ma, Q.G.\ Huang and X.\ Zhang,
\tia{Confronting brane inflation with Planck and pre-Planck data}
\doin{10.1103/PhysRevD.87.103516} {Phys.\ Rev.}{D}{87}{103516}{2013} [\arX{1303.6244}].
\bibitem{Suyama}
T.\ Suyama, T.\ Takahashi, M.\ Yamaguchi and S.\ Yokoyama,
\tia{Implications of Planck results for models with local type non-Gaussianity}  
\doij{10.1088/1475-7516/2013/06/012}{JCAP}{06}{012}{2013} [\arX{1303.5374}].
\bibitem{Naka}
K.\ Nakayama, F.\ Takahashi and T.T.\ Yanagida,
\tia{Polynomial Chaotic Inflation in the Planck Era}
\doin{10.1016/j.physletb.2013.06.050} {Phys.\ Lett.}{B}{725}{111}{2013} [\arX{1303.7315}].
\bibitem{Stein}
A.\ Ijjas, P.J.\ Steinhardt and A.\ Loeb,
\tia{Inflationary paradigm in trouble after Planck2013}
\doin{10.1016/j.physletb.2013.05.023} {Phys.\ Lett.}{B}{723}{261}{2013} [\arX{1304.2785}].
\bibitem{Kuro}  S.\ Tsujikawa, J.\ Ohashi, S.\ Kuroyanagi and A.\ De Felice, \tia{Planck constraints on single-field inflation} \doin{10.1103/PhysRevD.88.023529}{Phys.\ Rev.}{D}{88}{023529}{2013} 
[\arX{1305.3044}].
\bibitem{Burgess}
C.\ P.\ Burgess, M.\ Cicoli and F.\ Quevedo,
\tia{String Inflation After Planck 2013} \arX{1306.3512}.
\bibitem{Bar} 
S.\ Bartrum, M.\ Bastero-Gil, A.\ Berera, R.\ Cerezo, R.\ O.\ Ramos and J.\ G.\ Rosa,
\tia{The importance of being warm (during inflation)}  \arX{1307.5868}.

\bibitem{MaK}   R.\ Maartens and K.\ Koyama, \tia{Brane-world gravity} \doinn{10.12942/lrr-2010-5}{Living Rev.\ Rel.}{13}{5}{2010} [\arX{1004.3962}].
\bibitem{BH}    R.\ Brandenberger and P.-M.\ Ho, \tia{Noncommutative spacetime, stringy spacetime uncertainty principle, and density fluctuations}
\doin{10.1103/PhysRevD.66.023517}{Phys.\ Rev.}{D}{66}{023517}{2002} [\oarX{hep-th/0203119}].
\bibitem{RSa}   L.\ Randall and R.\ Sundrum, \tia{Large mass hierarchy from a small extra dimension} \doinn{10.1103/PhysRevLett.83.3370}{Phys.\ Rev.\ Lett.}{83}{3370}{1999} [\oarX{hep-ph/9905221}]. 
\bibitem{RSb}   L.\ Randall and R.\ Sundrum, \tia{Alternative to compactification} \doinn{10.1103/PhysRevLett.83.4690}{Phys.\ Rev.\ Lett.}{83}{4690}{1999} [\oarX{hep-th/9906064}].
\bibitem{KKL1}  J.E.\ Kim, B.\ Kyae and H.M.\ Lee, \tia{Effective Gauss--Bonnet interaction in Randall--Sundrum compactification} \doin{10.1103/PhysRevD.62.045013}{Phys.\ Rev.}{D}{62}{045013}{2000} [\oarX{hep-ph/9912344}].
\bibitem{KKL2}  J.E.\ Kim, B.\ Kyae and H.M.\ Lee, \tia{Various modified solutions of the Randall--Sundrum model with the Gauss--Bonnet interaction} \doin{10.1016/S0550-3213(00)00318-7}{Nucl.\ Phys.}{B}{582}{296}{2000} [\doin{10.1016/S0550-3213(00)00599-X}{Erratum-ibid.}{B}{591}{587}{2000}] [\oarX{hep-th/0004005}].
\bibitem{HL1}   Q.-G.\ Huang and M.\ Li, \tia{CMB power spectrum from noncommutative spacetime} \doij{10.1088/1126-6708/2003/06/014}{JHEP}{0306}{014}{2003} [\oarX{hep-th/0304203}].
\bibitem{Fukuma} M.\ Fukuma, Y.\ Kono and A.\ Miwa, \tia{Effects of space-time noncommutativity on the angular power spectrum of the CMB} \doin{10.1016/j.nuclphysb.2004.01.020}{Nucl.\ Phys.}{B}{682}{377}{2004} [\oarX{hep-th/0307029}].
\bibitem{TMB}    S.\ Tsujikawa, R.\ Maartens and R.\ Brandenberger, \tia{Noncommutative inflation and the CMB} \doin{10.1016/j.physletb.2003.09.022}{Phys.\ Lett.}{B}{574}{141}{2003} [\oarX{astro-ph/0308169}].
\bibitem{HL2}    Q.-G.\ Huang and M.\ Li, \tia{Noncommutative inflation and the CMB multipoles} \doij{10.1088/1475-7516/2003/11/001}{JCAP}{11}{001}{2003} [\oarX{astro-ph/0308458}].
\bibitem{HL3}    Q.-G.\ Huang and M.\ Li, \tia{Power spectra in spacetime noncommutative inflation} \doin{10.1016/j.nuclphysb.2005.02.002}{Nucl.\ Phys.}{B}{713}{219}{2005} [\oarX{astro-ph/0311378}].
\bibitem{KLM}    H.\ Kim, G.S.\ Lee and Y.S.\ Myung, \tia{Noncommutative spacetime effect on the slow-roll period of inflation} \doin{10.1142/S0217732305016518}{Mod.\ Phys.\ Lett.}{A}{20}{271}{2005} [\oarX{hep-th/0402018}].
\bibitem{KLLM}   H.\ Kim, G.S.\ Lee, H.W.\ Lee and Y.S.\ Myung, \tia{Second order corrections to noncommutative spacetime inflation} \doin{10.1103/PhysRevD.70.043521}{Phys.\ Rev.}{D}{70}{043521}{2004} [\oarX{hep-th/0402198}].
\bibitem{cai04}  R.-G.\ Cai, \tia{A Note on curvature fluctuation of noncommutative inflation} \doin{10.1016/j.physletb.2004.04.078}{Phys.\ Lett.}{B}{593}{1}{2004} [\oarX{hep-th/0403134}].
\bibitem{Cal4}   G.\ Calcagni, \tia{Noncommutative models in patch cosmology} \doin{10.1103/PhysRevD.70.103525}{Phys.\ Rev.}{D}{70}{103525}{2004} [\oarX{hep-th/0406006}].
\bibitem{Cal5}   G.\ Calcagni, \tia{Consistency relations and degeneracies in (non)commutative patch inflation} \doin{10.1016/j.physletb.2004.11.075}{Phys.\ Lett.}{B}{606}{177}{2005} [\oarX{hep-ph/0406057}].

\bibitem{CT04}  G.\ Calcagni and S.\ Tsujikawa, \tia{Observational constraints on patch inflation in noncommutative spacetime} \doin{10.1103/PhysRevD.70.103514}{Phys.\ Rev.}{D}{70}{103514}{2004} [\oarX{astro-ph/0407543}].

\bibitem{LOSW}  A.\ Lukas, B.A.\ Ovrut, K.S.\ Stelle and D.\ Waldram, \tia{The Universe as a domain wall} \doin{10.1103/PhysRevD.59.086001}{Phys.\ Rev.}{D}{59}{086001}{1999} [\oarX{hep-th/9803235}].
\bibitem{LOW1}  A.\ Lukas, B.A.\ Ovrut and D.\ Waldram, \tia{Cosmological solutions of Horava--Witten theory} \doin{10.1103/PhysRevD.60.086001}{Phys.\ Rev.}{D}{60}{086001}{1999} [\oarX{hep-th/9806022}].
\bibitem{BDL}   P.\ Bin\'{e}truy, C.\ Deffayet and D.\ Langlois, \tia{Nonconventional cosmology from a brane universe} \doin{10.1016/S0550-3213(99)00696-3}{Nucl.\ Phys.}{B}{565}{269}{2000} [\oarX{hep-th/9905012}].
\bibitem{BDEL}  P.\ Bin\'{e}truy, C.\ Deffayet, U.\ Ellwanger and D.\ Langlois, \tia{Brane cosmological evolution in a bulk with cosmological constant} \doin{10.1016/S0370-2693(00)00204-5}{Phys.\ Lett.}{B}{477}{285}{2000} [\oarX{hep-th/9910219}].
\bibitem{Shiromizu} T.\ Shiromizu, K.-i.\ Maeda and M.\ Sasaki, \tia{The Einstein equation on the 3-brane world} \doin{10.1103/PhysRevD.62.024012}{Phys.\ Rev.}{D}{62}{024012}{2000}
[\oarX{gr-qc/9910076}].
\bibitem{GrS}   D.J.\ Gross and J.H.\ Sloan, \tia{The quartic effective action for the heterotic string} \doin{10.1016/0550-3213(87)90465-2}{Nucl.\ Phys.}{B}{291}{41}{1987}.
\bibitem{CD}    C.\ Charmousis and J.-F.\ Dufaux, \tia{General Gauss--Bonnet brane cosmology} \doinn{10.1088/0264-9381/19/18/304}{Class.\ Quantum Grav.}{19}{4671}{2002} [\oarX{hep-th/0202107}].
\bibitem{dav03} S.C.\ Davis, \tia{Generalized Israel junction conditions for a Gauss--Bonnet brane world} \doin{10.1103/PhysRevD.67.024030}{Phys.\ Rev.}{D}{67}{024030}{2003} [\oarX{hep-th/0208205}].
\bibitem{GW}    E.\ Gravanis and S.\ Willison, \tia{Israel conditions for the Gauss--Bonnet theory and the Friedmann equation on the brane universe} \doin{10.1016/S0370-2693(03)00555-0}{Phys.\ Lett.}{B}{562}{118}{2003} [\oarX{hep-th/0209076}].
\bibitem{Torii}   K.\ Maeda and T.\ Torii,
\tia{Covariant gravitational equations on brane world with Gauss-Bonnet term}
\doin{10.1103/PhysRevD.69.024002}{Phys.\ Rev.}{D}{69}{024002}{2004} [\oarX{hep-th/0309152}].
\bibitem{Cal3}  G.\ Calcagni, \tia{Slow-roll parameters in braneworld cosmologies} \doin{10.1103/PhysRevD.69.103508}{Phys.\ Rev.}{D}{69}{103508}{2004} [\oarX{hep-ph/0402126}].
\bibitem{Sch} D.J.\ Schwarz, C.A.\ Terrero-Escalante and A.A.\ Garcia,
\tia{Higher order corrections to primordial spectra from cosmological inflation} 
\doin{10.1016/S0370-2693(01)01036-X}{Phys.\ Lett.}{B}{517}{243}{2001} [\oarX{astro-ph/0106020}].
\bibitem{Leach} S.M.\ Leach, A.R.\ Liddle, J.\ Martin and D.J.\ Schwarz,
\tia{Cosmological parameter estimation and the inflationary cosmology} 
 \doin{10.1103/PhysRevD.66.023515}{Phys.\ Rev.}{D}{66}{023515}{2002} [\oarX{astro-ph/0202094}].
\bibitem{Maar} R.\ Maartens, D.\ Wands, B.A.\ Bassett and I.\ Heard,
\tia{Chaotic inflation on the brane}
 \doin{10.1103/PhysRevD.62.041301}{Phys.\ Rev.}{D}{62}{041301}{2000} [\oarX{hep-ph/9912464}].
\bibitem{Langlois} D.\ Langlois, R.\ Maartens and D.\ Wands,
\tia{Gravitational waves from inflation on the brane} 
\doin{10.1016/S0370-2693(00)00957-6}{Phys.\ Lett.}{B}{489}{259}{2000} [\oarX{hep-th/0006007}]. 
\bibitem{Smith} A.R.\ Liddle and A.J.\ Smith,
\tia{Observational constraints on brane world chaotic inflation}
\doin{10.1103/PhysRevD.68.061301}{Phys.\ Rev.}{D}{68}{061301}{2003} [\oarX{astro-ph/0307017}]. 
\bibitem{TsuLiddle} S.\ Tsujikawa and A.R.\ Liddle,
\tia{Constraints on brane world inflation from CMB anisotropies}
\doij{10.1088/1475-7516/2004/03/001}{JCAP}{03}{001}{2004} [\oarX{astro-ph/0312162}]. 
\bibitem{Dufaux} J.F.\ Dufaux, J.E.\ Lidsey, R.\ Maartens and M.\ Sami,
\tia{Cosmological perturbations from brane inflation with a Gauss-Bonnet term}
 \doin{10.1103/PhysRevD.70.083525}{Phys.\ Rev.}{D}{70}{083525}{2004} [\oarX{hep-th/0404161}].
\bibitem{TSM} S.\ Tsujikawa, M.\ Sami and R.\ Maartens,   
\tia{Observational constraints on braneworld inflation: The Effect of a Gauss-Bonnet term}
\doin{10.1103/PhysRevD.70.063525 }{Phys.\ Rev.}{D}{70}{063525}{2004} [\oarX{astro-ph/0406078}].

\bibitem{yon87}  T.\ Yoneya, \tia{Duality and indeterminacy principle in string theory} in \textit{Wandering in the Fields}, K.\ Kawarabayashi and A.\ Ukawa (Eds.), World Scientific, Singapore (1987). 
\bibitem{LY}     M.\ Li and T.\ Yoneya, \tia{D particle dynamics and the space-time uncertainty relation} \doinn{10.1103/PhysRevLett.78.1219}{Phys.\ Rev.\ Lett.}{78}{1219}{1997} [\oarX{hep-th/9611072}].
\bibitem{yon00}  T.\ Yoneya, \tia{String theory and space-time uncertainty principle} \doinn{10.1143/PTP.103.1081}{Prog.\ Theor.\ Phys.}{103}{1081}{2000} [\oarX{hep-th/0004074}].

%
\bibitem{cosmomc} \url{http://cosmologist.info/cosmomc/}
\bibitem{Lewis} A.\ Lewis, \tia{Efficient sampling of fast and slow cosmological parameters} \doin{10.1103/PhysRevD.87.103529}{Phys.\ Rev.}{D}{87}{103529}{2013} [\arX{1304.4473}].
\bibitem{BAO} F.\ Beutler {\it et al.}, \tia{The 6dF Galaxy Survey: baryon acoustic oscillations and the local Hubble constant} \doinn{10.1111/j.1365-2966.2011.19250.x}{Mon.\ Not.\ Roy.\ Astron.\ Soc.}{416}{3017}{2011} [\arX{1106.3366}];\\
N.\ Padmanabhan {\it et al.}, \tia{A 2\% distance to $z=0.35$ by reconstructing baryon acoustic oscillations -- I: methods and application to the Sloan Digital Sky Survey}
\doinn{10.1111/j.1365-2966.2012.21888.x}{Mon.\ Not.\ Roy.\ Astron.\ Soc.}{427}{2132}{2012} [\arX{1202.0090}];\\
L.\ Anderson {\it et al.}, \tia{The clustering of galaxies in the SDSS-III Baryon Oscillation Spectroscopic Survey: baryon acoustic oscillations in the Data Release 9 Spectroscopic Galaxy Sample} \doinn{10.1111/j.1365-2966.2012.22066.x}{Mon.\ Not.\ Roy.\ Astron.\ Soc.}{427}{3435}{2013} [\arX{1203.6594}].
\bibitem{Das} S.\ Das {\it et al.}, \tia{The Atacama cosmology telescope: temperature and gravitational lensing power spectrum measurements from three seasons of data}
\arX{1301.1037};\\
C.L.\ Reichardt {\it et al.}, \tia{A measurement of secondary cosmic microwave background anisotropies with two years of South Pole Telescope observations}
\doinn{10.1088/0004-637X/755/1/70}{Astrophys.\ J.}{755}{70}{2012} [\arX{1111.0932}].
\bibitem{Malda} J.M.\ Maldacena, \tia{Non-Gaussian features of primordial fluctuations in single field inflationary models}
\doij{10.1088/1126-6708/2003/05/013}{JHEP}{05}{013}{2003} [\oarX{astro-ph/0210603}].
\bibitem{Cremi} P.\ Creminelli and M.\ Zaldarriaga, 
\tia{Single field consistency relation for the 3-point function}
\doij{10.1088/1475-7516/2004/10/006}{JCAP}{10}{006}{2004} [\oarX{astro-ph/0407059}].
\bibitem{TsuDe} A.\ De Felice and S.\ Tsujikawa, 
\tia{Shapes of primordial non-Gaussianities in the Horndeski's most general scalar-tensor theories}
\doij{10.1088/1475-7516/2013/03/030}{JCAP}{03}{030}{2013} [\arX{1301.5721}].
\bibitem{Planck3} P.A.R.\ Ade {\it et al.} [Planck Collaboration], \tia{Planck 2013 Results. XXIV. Constraints on primordial non-Gaussianity} \arX{1303.5084}.
\bibitem{Cal8} G.\ Calcagni, \tia{Non-Gaussianity in braneworld and tachyon inflation} \doij{10.1088/1475-7516/2005/10/009}{JCAP}{10}{009}{2005} [\oarX{astro-ph/0411773}].
\bibitem{MuM}   B.M.\ Murray and Y.S.\ Myung, \tia{Gauss--Bonnet braneworld and WMAP three year results} \doin{10.1016/j.physletb.2006.09.068}{Phys.\ Lett.}{B}{642}{426}{2006}
 [\oarX{astro-ph/0605684}].
\bibitem{BGS}   M.C.\ Bento, R.\ Gonz\'alez Felipe and N.M.C.\ Santos, \tia{Braneworld inflation from an effective field theory after WMAP three-year data}
  \doin{10.1103/PhysRevD.74.083503}{Phys.\ Rev.}{D}{74}{083503}{2006} [\oarX{astro-ph/0606047}].
\bibitem{BZCB}  A.\ Bouaouda, R.\ Zarrouki, H.\ Chakir and M.\ Bennai, \tia{$F$-term braneworld inflation in light of five-year WMAP observations}
  \doin{10.1142/S0217751X1004927X}{Int.\ J.\ Mod.\ Phys.}{A}{25}{3445}{2010} [\arX{1010.4884}].
\bibitem{NoR}   K.\ Nozari and N.\ Rashidi, \tia{Non-minimal braneworld inflation after Planck} \arX{1309.1950}.
\bibitem{LiZ13} N.\ Li and X. Zhang, \tia{Reexamination of inflation in noncommutative space-time after Planck results} \doin{10.1103/PhysRevD.88.023508}{Phys.\ Rev.}{D}{88}{023508}{2013} [\arX{1304.4358}].

\bibitem{Liddle:2003as} A.R.\ Liddle and S.M.\ Leach, \tia{How long before the end of inflation were observable perturbations produced?} \doin{10.1103/PhysRevD.68.103503}{Phys.\ Rev.}{D}{68}{103503}{2003} [\oarX{astro-ph/0305263}].
\bibitem{Copeland:2005qe} E.J.\ Copeland and O.\ Seto, \tia{Reheating and gravitino production in braneworld inflation} \doin{10.1103/PhysRevD.72.023506}{Phys.\ Rev.}{D}{72}{023506}{2005} [\oarX{hep-ph/0505149}].

\bibitem{chaotic} A.D.\ Linde, \tia{Chaotic inflation} \doin{10.1016/0370-2693(83)90837-7}{Phys.\ Lett.}{B}{129}{177}{1983}.
\bibitem{linear}  E.\ Silverstein and A.\ Westphal, 
\tia{Monodromy in the CMB: gravity waves and string inflation}
\doin{10.1103/PhysRevD.78.106003}{Phys.\ Rev.}{D}{78}{106003}{2008} [\arX{0803.3085}];\\
L.\ McAllister, E.\ Silverstein and A.\ Westphal, 
\tia{Gravity waves and linear inflation from axion monodromy}
\doin{10.1103/PhysRevD.82.046003}{Phys.\ Rev.}{D}{82}{046003}{2010} [\arX{0808.0706}].
\bibitem{exponential} F.\ Lucchin and S.\ Matarrese, \tia{Power-law inflation} \doin{10.1103/PhysRevD.32.1316}{Phys.\ Rev.}{D}{32}{1316}{1985};\\
J.J.\ Halliwell, \tia{Scalar fields in cosmology with an exponential potential} \doin{10.1016/0370-2693(87)91011-2}{Phys.\ Lett.}{B}{185}{341}{1987};\\
Y.\ Kitada and K.-i.\ Maeda, \tia{Cosmic no-hair theorem in power-law inflation} \doin{10.1103/PhysRevD.45.1416}{Phys.\ Rev.}{D}{45}{1416}{1992}.
\bibitem{natural} K.\ Freese, J.A.\ Frieman and A.V.\ Olinto, \tia{Natural inflation with pseudo Nambu--Goldstone bosons} \doin{10.1103/PhysRevLett.65.3233}{Phys.\ Rev.\ Lett.}{}{65}{3233}{1990};\\
F.C.\ Adams, J.R.\ Bond, K.\ Freese, J.A.\ Frieman and A.V.\ Olinto, \tia{Natural inflation: particle physics models, power-law spectra for large-scale structure, and constraints from COBE} \doin{10.1103/PhysRevD.47.426}{Phys.\ Rev.}{D}{47}{426}{1993} [\oarX{hep-ph/9207245}].
\bibitem{Star} A.A.\ Starobinsky, \tia{A new type of isotropic cosmological models without singularity} \doin{10.1016/0370-2693(80)90670-X}{Phys.\ Lett.}{B}{91}{99}{1980}.
\bibitem{Maeda} K.-i.\ Maeda, \tia{Towards the Einstein--Hilbert action via conformal transformation} \doin{10.1103/PhysRevD.39.3159}{Phys.\ Rev.}{D}{39}{3159}{1989};\\
A.\ De Felice and S.\ Tsujikawa, \tia{$f(R)$ theories} \doinn{10.12942/lrr-2010-3}{Living Rev.\ Rel.}{13}{3}{2010} [\arX{1002.4928}].

\bibitem{MSSM}  R.\ Allahverdi, K.\ Enqvist, J.\ Garc\'ia-Bellido and A.\ Mazumdar, \tia{Gauge invariant MSSM inflaton} \doinn{10.1103/PhysRevLett.97.191304}{Phys.\ Rev.\ Lett.}{97}{191304}{2006} [\oarX{hep-ph/0605035}]. 
\bibitem{infection}  R.\ Allahverdi, B.\ Dutta and A.\ Mazumdar,
\tia{Unifying inflation and dark matter with neutrino masses} \doinn{10.1103/PhysRevLett.99.261301}{Phys.\ Rev.\ Lett.}{97}{261301}{2007} [\arX{0708.3983}]. 
\bibitem{tip} L.\ Lorenz, J.\ Martin and C.\ Ringeval, \tia{Brane inflation and the WMAP data: a Bayesian analysis} \doij{10.1088/1475-7516/2008/04/001}{JCAP}{08}{04}{2008} [\arX{0709.37584}].
\end{thebibliography}
\end{document}